\documentclass[conference]{IEEEtran}
\IEEEoverridecommandlockouts

\usepackage{booktabs}
\usepackage{lipsum} 
\usepackage{cite}
\usepackage{amsmath,amssymb,amsfonts}
\usepackage{algorithmic}
\usepackage{graphicx}
\usepackage{textcomp}
\usepackage[table]{xcolor}
\usepackage{multirow}

\usepackage{tcolorbox}
\usepackage{dblfloatfix}
\usepackage{xspace}
\usepackage{url}

\usepackage{tikz}
\def\checkmark{\tikz\fill[scale=0.4](0,.35) -- (.25,0) -- (1,.7) -- (.25,.15) -- cycle;} 

\definecolor{lightgray}{gray}{0.9}

\newcommand{\eg}{\emph{e.g.,}\xspace}
\newcommand{\ie}{\emph{i.e.,}\xspace}

\begin{document}

\title{Framework Code Samples: How Are They Maintained and Used by Developers? }

\author{\IEEEauthorblockN{Gabriel Menezes\IEEEauthorrefmark{1}, Bruno Cafeo\IEEEauthorrefmark{1}, Andre Hora\IEEEauthorrefmark{2}}
\IEEEauthorblockA{\IEEEauthorrefmark{1}
    Faculty of Computing, UFMS, Brazil \\
menezes@aluno.ufms.br, cafeo@facom.ufms.br}
\IEEEauthorblockA{\IEEEauthorrefmark{2}
    Department of Computer Science, UFMG, Brazil \\
 andrehora@dcc.ufmg.br}
}

\IEEEoverridecommandlockouts
\IEEEpubid{\makebox[\columnwidth]{978-1-7281-2968-6/19/\$31.00~\copyright2019 IEEE \hfill} \hspace{\columnsep}\makebox[\columnwidth]{ }}

\maketitle

\IEEEpubidadjcol

\begin{abstract}
\emph{Background:} Modern software systems are commonly built on the top of frameworks.
To accelerate the learning process of features provided by frameworks, code samples are made available to assist developers.
However, we know little about how code samples are actually developed.
\emph{Aims:} In this paper, we aim to fill this gap by assessing the characteristics of framework code samples.
We provide insights on how code samples are maintained and used by developers.
\emph{Method:} We analyze 233 code samples of Android and SpringBoot, and assess aspects related to their source code, evolution, popularity, and client usage.
\emph{Results:} We find that most code samples are small and simple, provide a working environment to the clients, and rely on automated build tools.
They change frequently over time, for example, to adapt to new framework versions.
We also detect that clients commonly fork the code samples, however, they rarely modify them.
\emph{Conclusions:} We provide a set of lessons learned and implications to creators and clients of code samples to improve maintenance and usage activities.
\end{abstract}


\section{Introduction}
\label{sec:introduction}

Modern software systems are commonly implemented with the support of frameworks, which provide feature reuse, improve productivity, and decrease costs~\cite{Mose96, Kons09, raemaekers12}.
Frameworks support the development of mobile apps, web platforms, responsive interfaces, cross-platform systems, among other.
In the Java ecosystem, for example, there are more than 270,000 packages available to be used by client systems in the Maven repository.\footnote{\url{https://search.maven.org/stats}}
In the JavaScript ecosystem the numbers are even higher: the npm repository has over 400,000 packages and reports 6 billions downloads in a single month.\footnote{https://www.linux.com/news/event/Nodejs/2016/state-union-npm}

To facilitate and accelerate the learning process of features provided by frameworks, code samples are commonly made available to assist development efforts~\cite{mozilla18}.
Code samples are often provided by world-wide software projects and organizations, such as Android,\footnote{\url{https://developer.android.com/samples}} Spring,\footnote{\url{https://spring.io/guides}}
Google Maps,\footnote{\url{https://developers.google.com/maps/documentation/javascript/examples}}
Twitter,\footnote{\url{http://twitterdev.github.io}} 
Microsoft,\footnote{\url{https://code.msdn.microsoft.com}} to name a few.
Framework code samples may introduce the usage of basic features, as well as more advanced ones.
For instance, a basic sample provided by the SpringBoot framework help newcomer developers on building RESTful web services.\footnote{\url{https://spring.io/guides/gs/rest-service}}
In contrast, a more advanced code sample made available by the same framework help developers on securing web applications.\footnote{\url{https://spring.io/guides/gs/securing-web}}
Due to their practicality, client developers may copy and paste code samples into their own code base, and may put them into production~\cite{mozilla18}.
Thus, ideally, code samples should follow some good development practices, such as be simple, small, self-contained, easy to understand, secure, and efficient~\cite{mozilla18}.

Although framework code samples are commonly available to help developers, we know little about how they are actually maintained and used by developers.
In this context, some questions are still opened, such as:
what is the common size of code samples?
how do code samples evolve over time?
what makes a code sample more popular than other? 
how are the code samples used by the developers?
By answering these questions, we can assess common aspects of code samples, better supporting their maintenance and usage activities.

In this paper, we aim to fill this gap by assessing the characteristics of framework code samples.
Specifically, we analyze 233 code samples provided by two widely popular frameworks: Android and SpringBoot.
We answer four research questions related to their maintenance and usage.
Particularly, we assess aspects related to their source code, evolution, popularity, and client usage:

\begin{itemize}

\item \emph{RQ1 (Source Code):
What are the source code characteristics of framework code samples?}
We find that framework code samples are overall simple and small.
We also detect that code samples rely on automated build tools and provide working environments to facilitate the task of running them.

\item \emph{RQ2 (Evolution): 
How do framework code samples evolve over time?}
We detect that code samples are not static, but they evolve over time.
Updates are often made to keep them up to date with new framework versions, and, consequently, relevant to the clients.

\item \emph{RQ3 (Popularity): Which aspects differentiate popular framework samples from ordinary ones?}
By comparing popular and unpopular code samples, we find that the popular ones are more likely to have a higher amount of source code files. They are also more likely to change over time than the unpopular ones.

\item \emph{RQ4 (Client Usage): How are the framework code samples used by developers?}
We rely on the fork metric as a proxy of code sample usage.
We find that the majority of the forked code samples are inactive.
However, a non-negligible ratio of the forked code samples are updated.


\end{itemize}

\smallskip

Based on our findings, we provide a set of implications to code sample creators and clients, particularly, to support their maintenance and usage activities.

\smallskip

\noindent\emph{Contributions:} 
This paper has three major contributions:

\begin{enumerate}

\item To the best of our knowledge, this is the first research to assess framework code samples, which support the learning process of features provided by frameworks.

\item We provide a large empirical study on the code samples made available by Android and SpringBoot to better understand their maintenance and usage practices.

\item We provide a set of lessons learned and implications to code sample creators and clients.

\end{enumerate}

\smallskip

\noindent\emph{Structure of the paper:} 
Section~\ref{sec:sample} introduces code samples and their importance to support development nowadays.
Section~\ref{sec:design} presents the study design, while Section~\ref{sec:results} reports the results.
Section~\ref{sec:summary} discusses the implications and Section~\ref{sec:threats} presents the threats to validity.
Finally, Section~\ref{sec:related} discusses related work and Section~\ref{sec:conclu} concludes the paper.

\section{Code Samples in a Nutshell}
\label{sec:sample}

Framework code samples aim to facilitate and accelerate the learning process of features provided by frameworks.
In this context, Oracle states that ``\emph{code sample is provided for educational purposes or to assist your development or administration efforts}''.\footnote{\url{https://www.oracle.com/technetwork/indexes/samplecode}}
Spring reports that ``\emph{code samples are designed to get you productive as quickly as possible}''.\footnote{\url{https://spring.io/guides}}

Popular frameworks make code samples available to assist their client developers.
The Android framework, for example, has more than one hundred samples on GitHub to help the creation of mobile apps.
The SpringBoot framework also has dozens of samples to support the implementation of web apps. 
In addition to those well-known frameworks, code samples are often provided by organizations to facilitate the usage of their technologies, such as Google Maps APIs, Twitter APIs, Microsoft platforms, Apple platforms, among other.

In order to create good code samples, some guidelines are available.
For example, the \emph{Code example guidelines} provided by Mozilla~\cite{mozilla18} states general practices related to the size, understandability, simplicity, self-containment, security, and efficiency.
Guidelines also exist to set up the formatting of code samples, as the one provided by Google.\footnote{\url{https://developers.google.com/style/code-samples}}
In addition, numerous blogs on programming practices support the developers who are in charge of creating code samples.\footnote{\eg \url{https://goo.gl/SzV5PL}, \url{https://goo.gl/QaA16L}, \url{https://goo.gl/ixGaqF}}

\begin{figure}[h]
    \centering
    \includegraphics[width=0.47\textwidth]{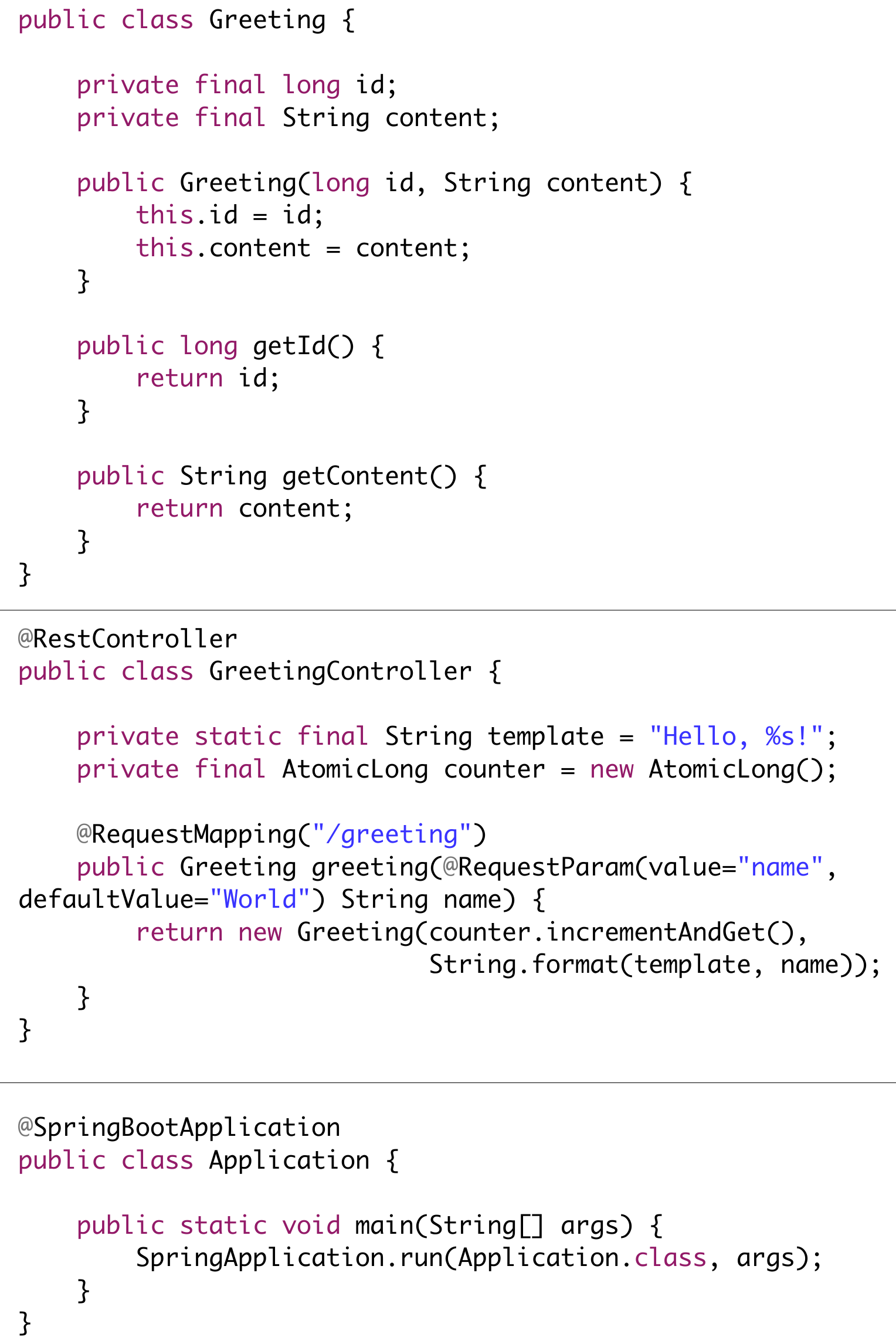}
    \caption{Example of code sample (SpringBoot framework).}
    \label{fig:example1}
\end{figure}

Figure~\ref{fig:example1} presents an official code sample provided by the SpringBoot framework.\footnote{\url{https://spring.io/guides/gs/rest-service}}
It supports new developers on building RESTful web services.
This sample is composed by only three major classes and 
helps the clients dealing with important SpringBoot features provided via the annotations: \texttt{@Rest\-Controller}, \texttt{@Request\-Mapping}, \texttt{@Request\-Param}, and \texttt{@Spring\-Boot\-Application}.
This sample is also composed by other files (\eg~xml, json, shell) to properly help the client running it.

\begin{figure}[h]
    \centering
    \includegraphics[width=0.48\textwidth]{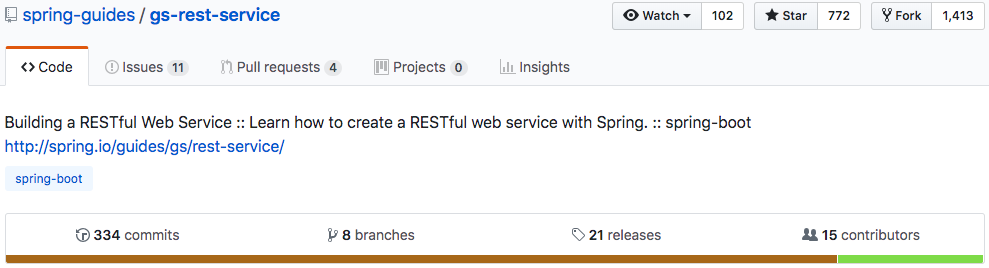}
    \quad
    \centering
    \includegraphics[width=0.48\textwidth]{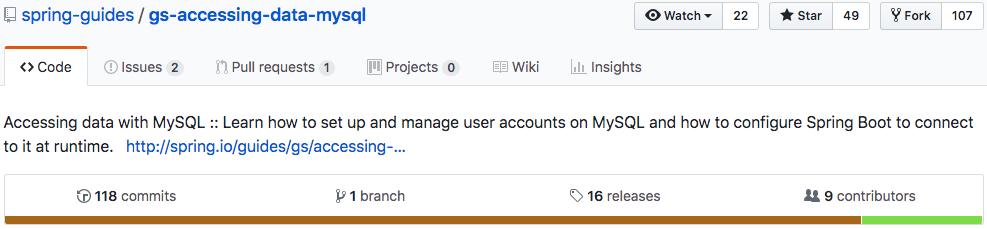}
    \caption{Code sample statistics (SpringBoot framework).}
    \label{fig:example1-github}
\end{figure}

Although simple and small, the GitHub project\footnote{https://github.com/spring-guides/gs-rest-service} hosting this sample has 772 stars and 1,413 forks\footnote{A fork is a copy of a repository. It allows developers to change the copy without affecting the original project.}, suggesting that it is indeed relevant and helpful for developers, as presented in Figure~\ref{fig:example1-github} (top).
Interestingly, this sample is an active project: the 334 commits show that it is evolving over time.
By checking its changes, we notice many of them are made to update documentation and configuration files.
Changes are also performed to migrate the sample to new framework versions, keeping it up to date and ready to be used with fresh releases of SpringBoot.
However, not all code samples receive the same attention from the developers: another official sample provided by SpringBoot to access data with MySQL\footnote{https://github.com/spring-guides/gs-accessing-data-mysql} is much less popular (49 stars), grab less attention from the community (107 forks), and is less active (118 commits), as shown in Figure~\ref{fig:example1-github} (bottom).

Overall, we notice the relevance of code samples to support development, as exposed by the variety of software technologies that make them available.
We also verify the concerns to create good code samples, as pointed by the many available guidelines.
Finally, we notice that code samples can have distinct different levels of popularity, activity, and community engagement.

\smallskip

\begin{tcolorbox}[left=0mm,right=0mm,boxrule=0.1mm,colback=gray!30!white]
\vspace{-0.1cm}
Despite their importance, to the best of our knowledge, framework code samples are understudied.
We are not aware about fundamental aspects on how they are actually maintained and used by developers.
By reveling these aspects, we aim to better understand the code samples and provide initial insights on their maintenance and usage practices.
\vspace{-0.1cm}
\end{tcolorbox}

\section{Study Design}
\label{sec:design}

\subsection{Selecting the Case Studies}

In this study, we assess the code samples provided by two world wide frameworks: Android and SpringBoot.

The Android framework\footnote{\url{https://developer.android.com}} allows the creation of Android apps for several devices, such as smartphones, smartwatches, and TVs.
Android code samples are publicly available on GitHub\footnote{\url{https://github.com/googlesamples}} and help developers dealing with Android features, such as permissions, picture and video manipulation, background tasks, notifications, networks, multiple touch events, among many other.
The SpringBoot framework\footnote{\url{https://spring.io}} mostly support the development of web applications.
It also provides a set of code samples publicly available on GitHub\footnote{\url{https://github.com/spring-guides}} to help developers creating web apps, such as dealing with RESTful web services, scheduling tasks, uploading files, validating form inputs, caching data, securing apps, among other.
Considering both frameworks, in this study we analyze 233 code samples: 176 from Android and 57 from SpringBoot.

We select these two frameworks due to several reasons.
\emph{First}, they are relevant and worldwide adopted frameworks that have millions of users.
\emph{Second}, they support the creation of two distinct and important niche of apps: mobile and web.
\emph{Third}, their base of code samples are publicly available on GitHub, thus, in addition to access their source code, we can also perform evolutionary analysis.
\emph{Fourth}, they have a large base of developers, so we can better assess their usage.

\begin{figure}[h]
    \centering
    \includegraphics[width=0.155\textwidth]{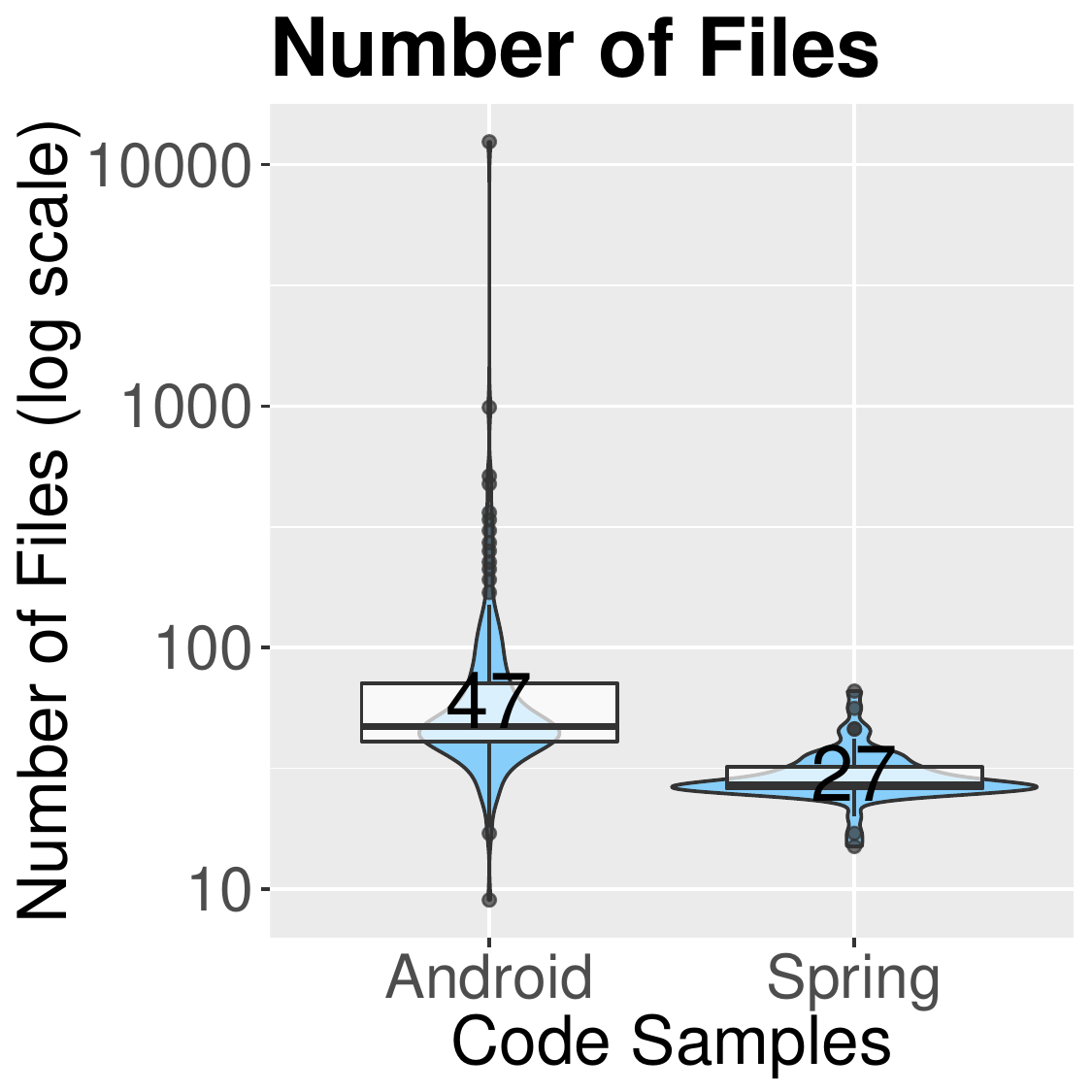}
    \includegraphics[width=0.155\textwidth]{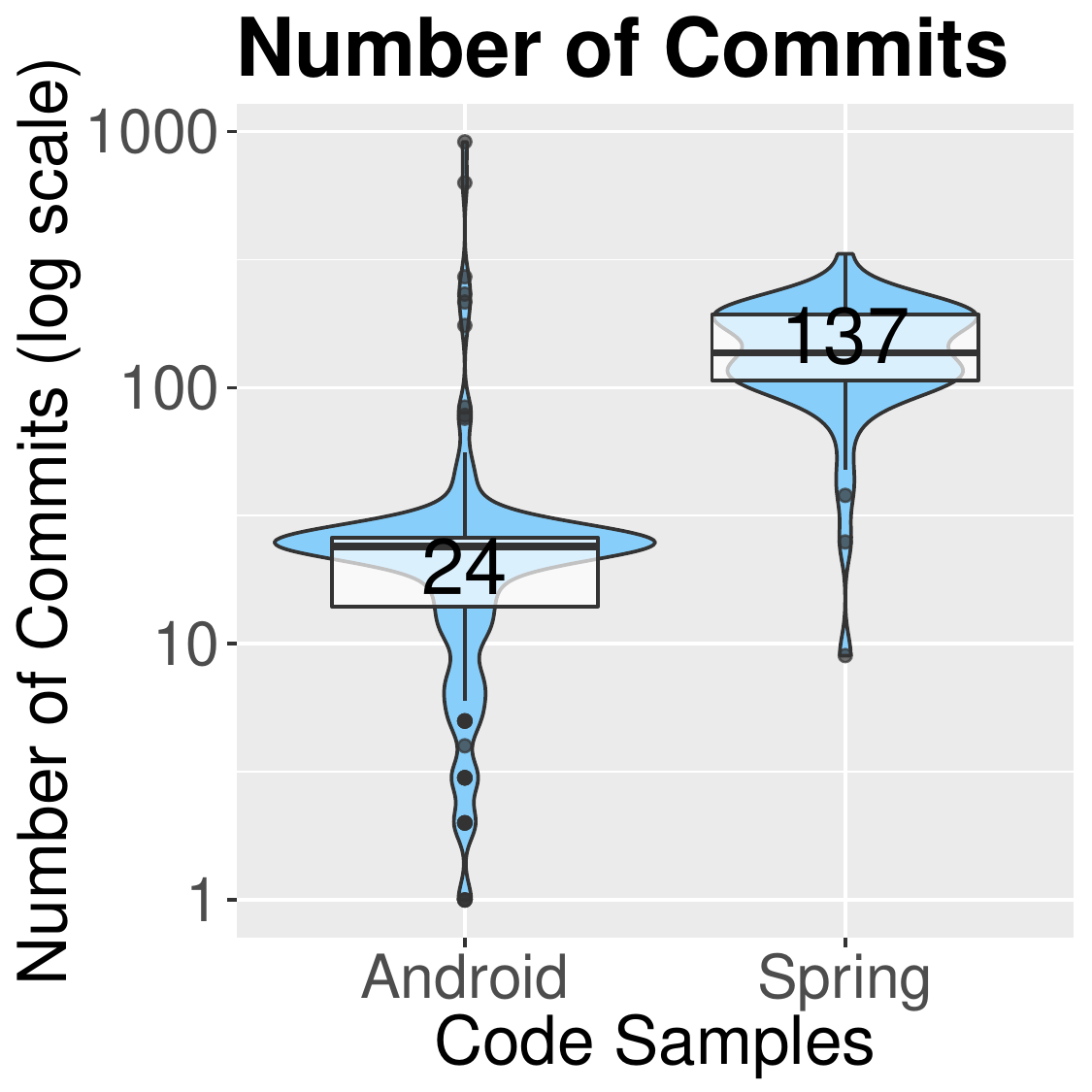}
    \includegraphics[width=0.155\textwidth]{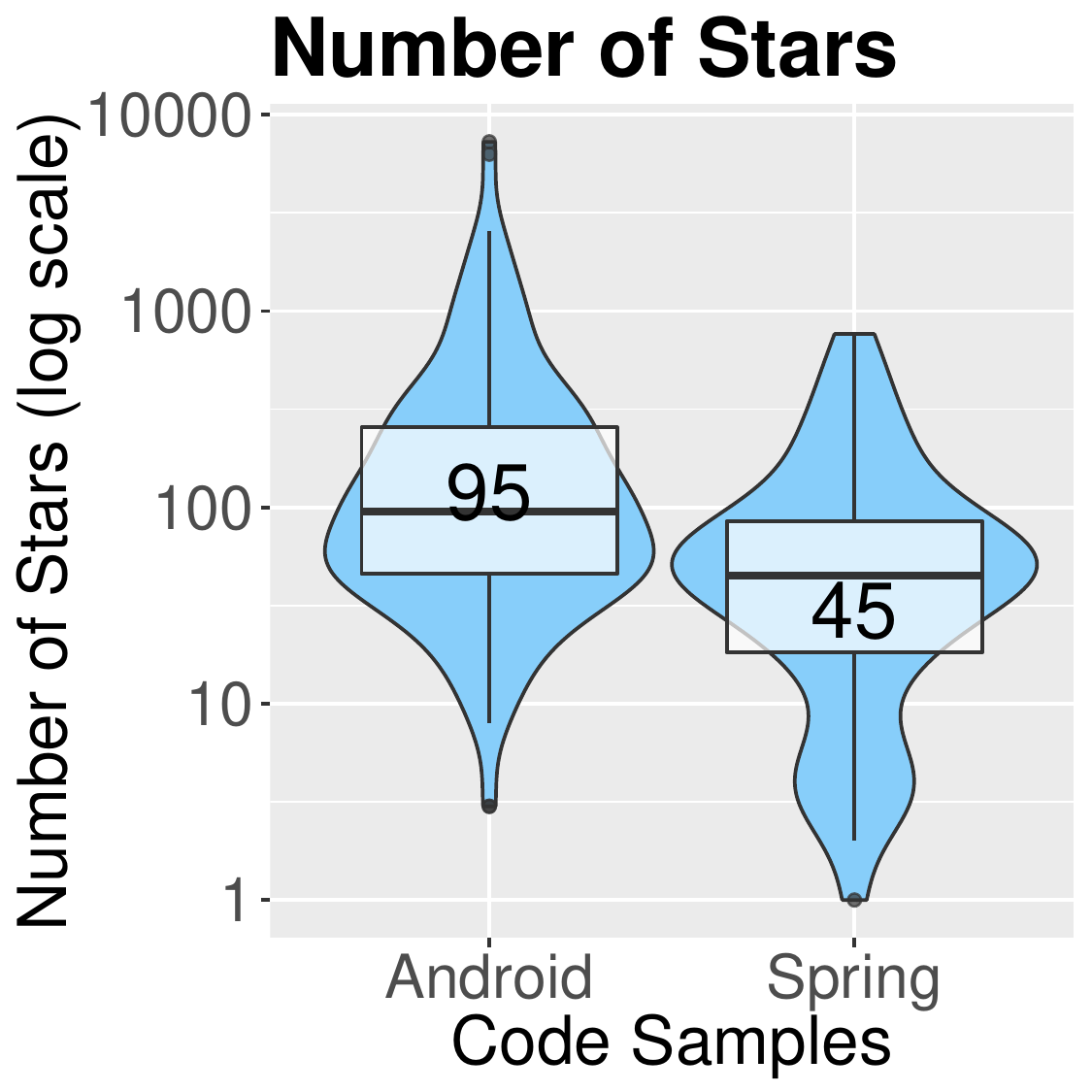}

    \caption{Basic metrics of the Android and SpringBoot code samples.}
    \label{fig:stats}
\end{figure}

Figure~\ref{fig:stats} presents the distribution of number of files, commits, and stars for the 233 code samples.
On the median, the Android code samples have 47 files, 24 commits, and 95 stars.
The most popular Android sample is \texttt{easy\-permissions} with 7,328 stars.
The SpringBoot code samples have 27 files, 137 commits, and 45 stars.
In this case, the most popular sample is \texttt{gs-rest-service} with 772 stars.

\subsection{Source Code Analysis (RQ1)}

\begin{figure}[h]
    \centering
    \includegraphics[width=0.47\textwidth]{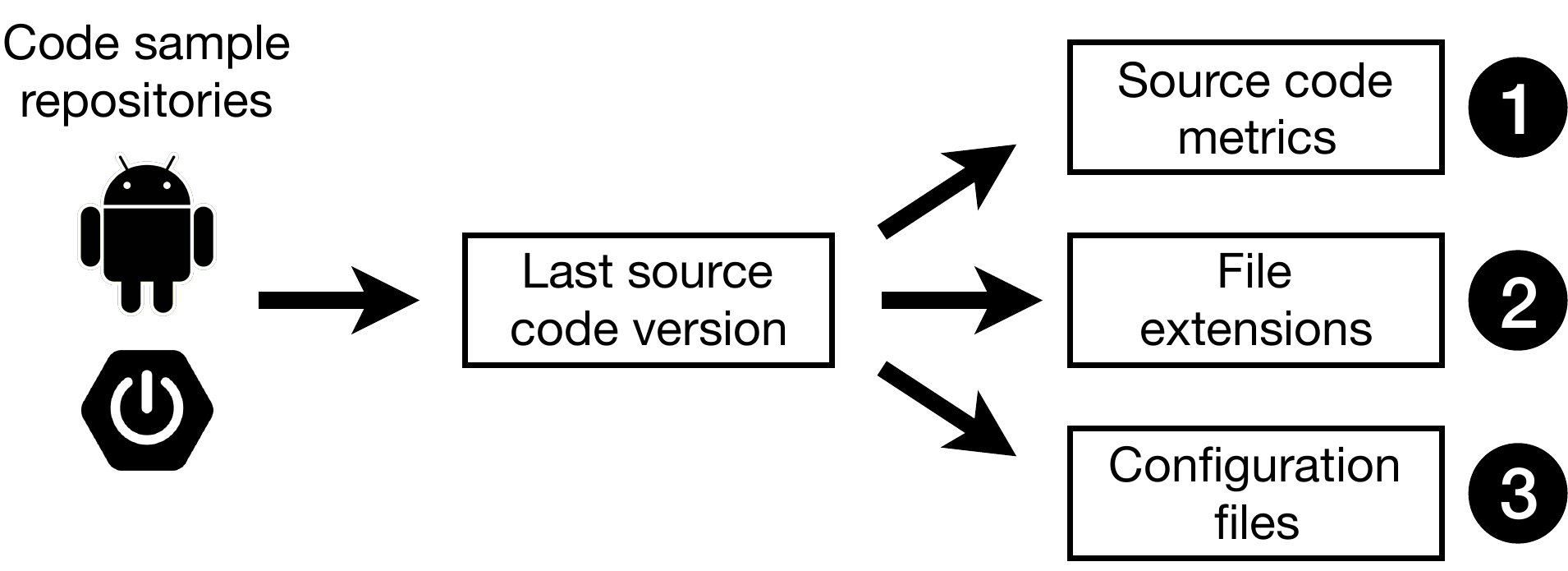}
    \caption{Source code analysis (RQ1).}
    \label{fig:rq1-steps}
\end{figure}

In Research Question 1, we assess the last version of the source code samples and extract three data: source code metrics, file extensions, and configuration files, as summarized in Figure~\ref{fig:rq1-steps}.

\smallskip

\noindent\emph{1. Source code metrics:}
We first assess the current state of the samples by computing source code metrics with the support of the software analysis tool Understand.\footnote{https://scitools.com}
Particularly, we focus on four metrics: number of java files, lines of code, cyclomatic complexity, and commented code lines.
\underline{Rationale}:
Small code with simple structures may improve code understanding and readability~\cite{martin2009clean}.
Code samples are not different; ideally, they should be concise and simple~\cite{mozilla18}.
Code comment is important to any piece of code~\cite{lethbridge2003software}, however, it may be even more relevant to samples as they provide inline comments to help the users.

\smallskip

\noindent\emph{2. File extensions:}
We extract the file extensions found in the code samples to better understand their content in addition to source code files.
\underline{Rationale:} In addition to java files, we are not aware of the files that are present in the code samples.
The higher presence of other files (\eg~xml, json, jars etc) may indicate that a working environment is available to the clients to run the code samples.
In contrast, if the files are mostly concentrated on Java, this may suggest that additional work is still needed by the clients to properly set up the environment.

\smallskip

\noindent\emph{3. Configuration files:}
In addition to the file extensions, we also compute the most common configuration files from the code samples.
Particularly, we verify whether the code samples adopts automation tools to build, integrate, and manage dependency.
\underline{Rationale:} By relying on these automation tools, the framework code samples are following good development practices, which are commonly adopted on software projects to improve quality and productivity and reduce risks~\cite{DuvallMatyasGlover07, Meyer2014, Vasilescu2015}.

\subsection{Evolutionary Analysis (RQ2)}

\begin{figure}[h]
    \centering
    \includegraphics[width=0.47\textwidth]{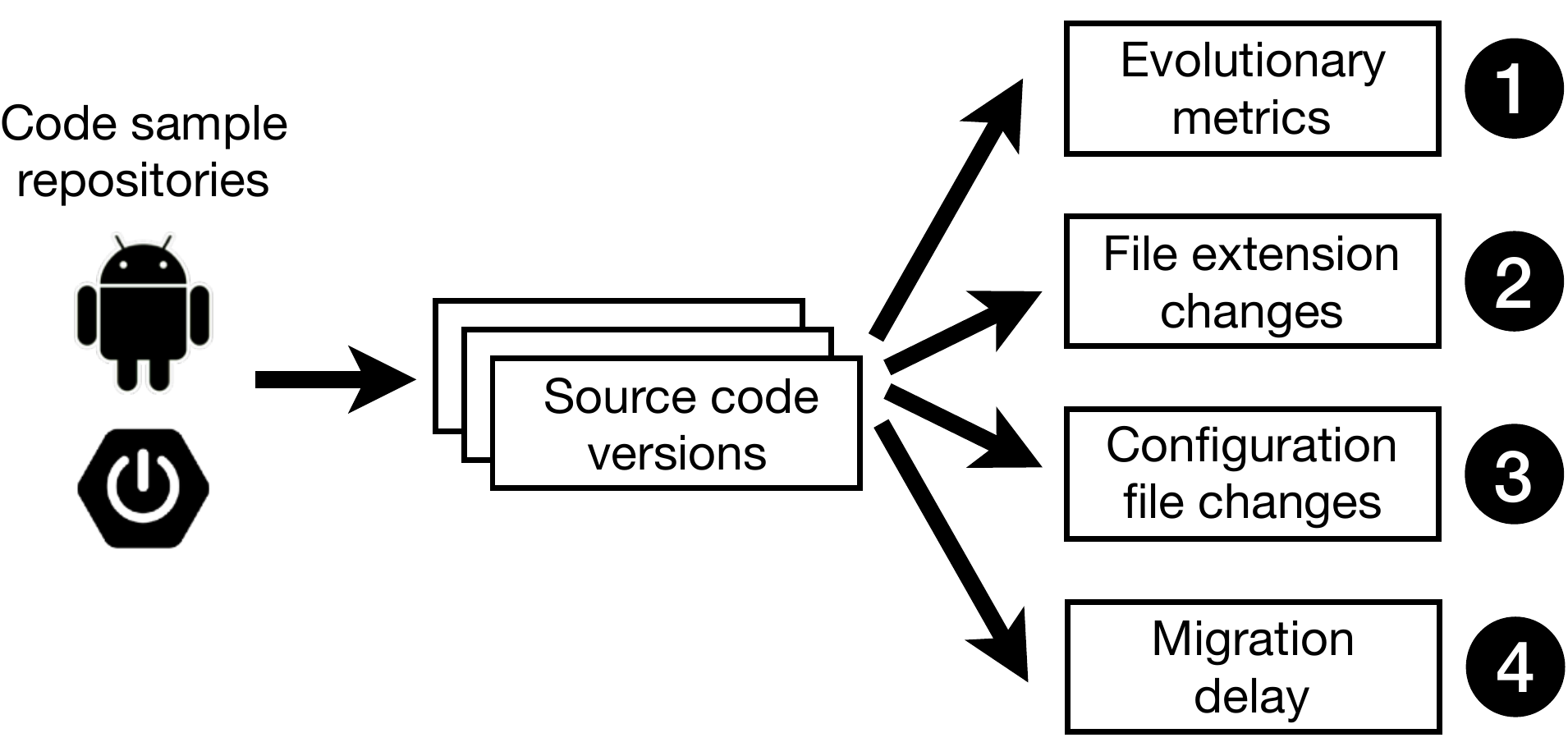}
    \caption{Evolutionary analysis (RQ2).}
    \label{fig:rq2-steps}
\end{figure}

In Research Question 2, we assess all the versions (\ie~commits) of the code samples and extract: evolutionary metrics, file extension changes, configuration file changes, and migration delay, as presented in Figure~\ref{fig:rq2-steps}.

\smallskip

\noindent\emph{1. Evolutionary metrics:}
We compute metrics to assess the evolution of the code samples.
Specifically, we extract two evolutionary metrics: frequency of commits and lifetime.
Lifetime is computed as the number of days between the first and the last project commit.
\underline{Rationale}:
To cope with API evolution~\cite{dig2006apis, laerte17, kula2018empirical, icse2018}, ideally, the code samples should change over time.
Code samples with frequent changes may indicate efforts to keep them up to date.
In contrast, less active code samples may suggest they are abandoned.

\smallskip

\noindent\emph{2. File extension changes:}
We analyze the file extension changes over time to better understand how the code samples are actually maintained.
\underline{Rationale}:
To evolve the code samples, source code and other files should be updated.
However, we are not aware of which files are most relevant to keep the samples properly working.

\vspace{0.5cm}

\noindent\emph{3. Configuration file changes:}
We analyze the modifications in the configuration files to assess whether the automation tools are being updated.
\underline{Rationale}:
In addition to use automation tools to build, integrate, and manage dependencies, it is important to keep them alive, otherwise, the advantages provided by these tools are not achieved.

\smallskip

\noindent\emph{4. Migration delay:}
We compute the migration delay between code samples and their frameworks.
In other words, we assess how long it takes for code samples migrate to new framework versions.
\underline{Rationale}:
As client projects, code samples are dependent of their frameworks.
When these frameworks evolve and provide new versions, the code samples (as any other framework client software) should be updated, otherwise, they will be frozen on past versions, and become less attractive to their users~\cite{McDo13, Robb12, laerte17, sqj2018, Kula2018}.

\subsection{Popularity Analysis (RQ3)}

\begin{figure}[h]
    \centering
    \includegraphics[width=0.49\textwidth]{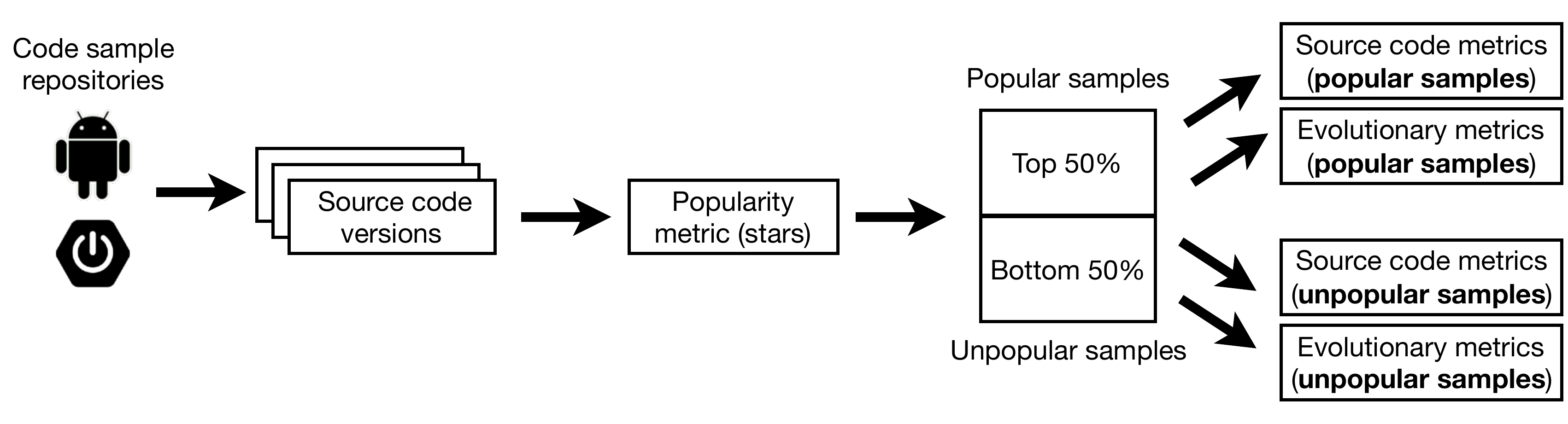}
    \caption{Popularity analysis (RQ3).}
    \label{fig:rq3-steps}
\end{figure}

In Research Question 3, we analyze the popularity of the studied code samples to find differences between the most and least popular.
Specifically, we sort the code samples in descending order according to their popularity in number of stars.
We classify as popular code samples the top 50\% with the highest number of stars.
Similarly, we classify as unpopular code samples the bottom 50\% with the lowest number of stars.
We then compare each group regarding the source code and evolutionary metrics described in RQs 1 and 2 (\eg~lines of code, complexity, lifetime, etc), as summarized in Figure~\ref{fig:rq3-steps}.
We also analyze the statistical significance of the difference between the groups by applying the Mann-Whitney test at \emph{p-value} = 0.05. 
To show the effect size of the difference between them, we compute Cliff's Delta (or \emph{d}); we use the \emph{effsize} package in R\footnote{https://cran.r-project.org/web/packages/effsize} to compute Cliff's Delta.
Following previous guidelines~\cite{romano2006appropriate}, we interpret the effect size values as negligible for $d<0.147$, small for $d < 0.33$, medium for 
$d < 0.474$, and large otherwise.

\smallskip

\noindent\underline{Rationale:} Several previous studies have used a similar approach to find differences between popular and unpopular software artifacts, for example, by assessing the popularity of mobile apps~\cite{tian15} and software libraries~\cite{laerte17,Brito18}.
Here, we adopt a similar approach to differentiate popular and unpopular code samples, learning with the practices provided by the popular ones.

\vspace{0.3cm}

\subsection{Client Usage Analysis (RQ4)}

\begin{figure}[h]
    \centering
    \includegraphics[width=0.47\textwidth]{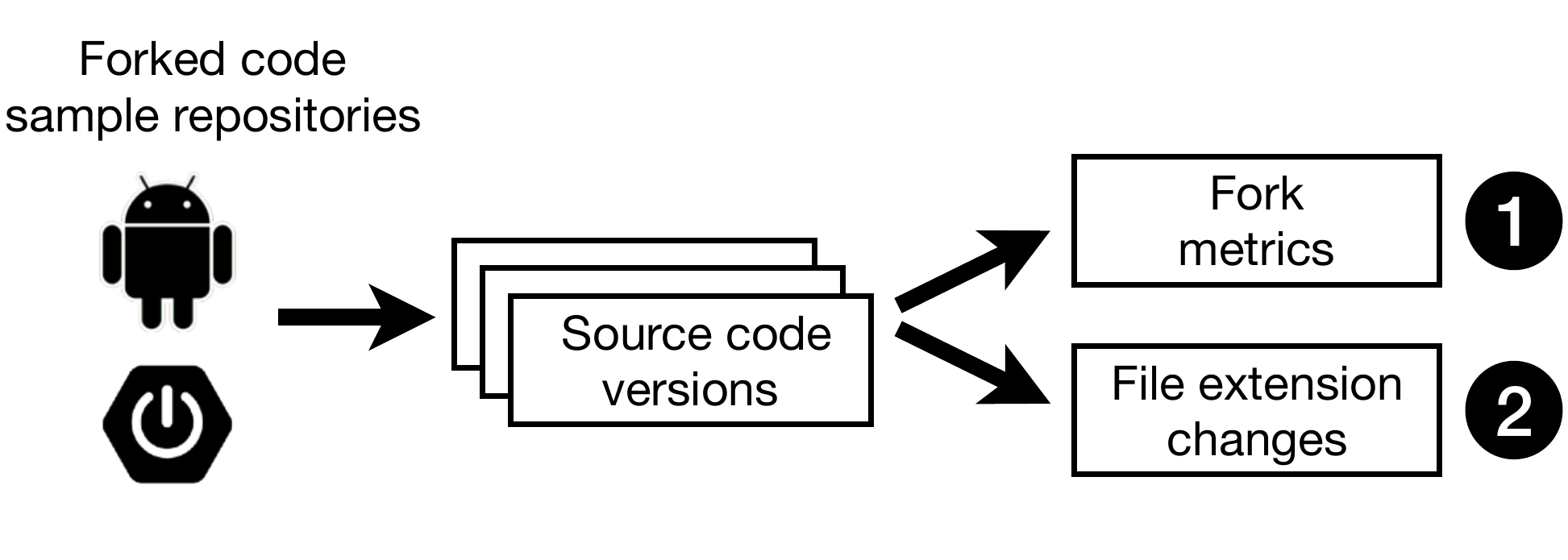}
    \caption{Client usage analysis (RQ4).}
    \label{fig:rq4-steps}
\end{figure}

In our last Research Question, we focus on the client side, that is, the developers who are adopting the code samples.
Particularly, we analyze all GitHub projects that forked the official code samples and compute: fork metrics and file extension changes, as summarized in Figure~\ref{fig:rq4-steps}.

\smallskip

\noindent\emph{1. Fork metrics:}
We compute three metrics to assess how the code samples are forked: number of forks, number of forks with commits, and number of commits in forked code samples.
\underline{Rationale}: Fork can be seen as a measure of popularity~\cite{hudson16}.
After forking, the client developer can update the code or simply do not perform any change.
In the case the forked project is updated, this may indicate that the client developer is somehow exploring the code sample, possibly, by running and improving it.

\smallskip

\noindent\emph{2. File extension changes:}
We also analyze the file extension changes to better understand how the forked code samples are actually updated.
\underline{Rationale}:
To evolve the forked code samples, source code and other files should be updated.
However, we are not aware which files are most relevant to be explored by the clients.

\section{Results}
\label{sec:results}

\subsection{Source Code (RQ1)}

\begin{figure}[h]
    \centering
    \includegraphics[width=0.23\textwidth]{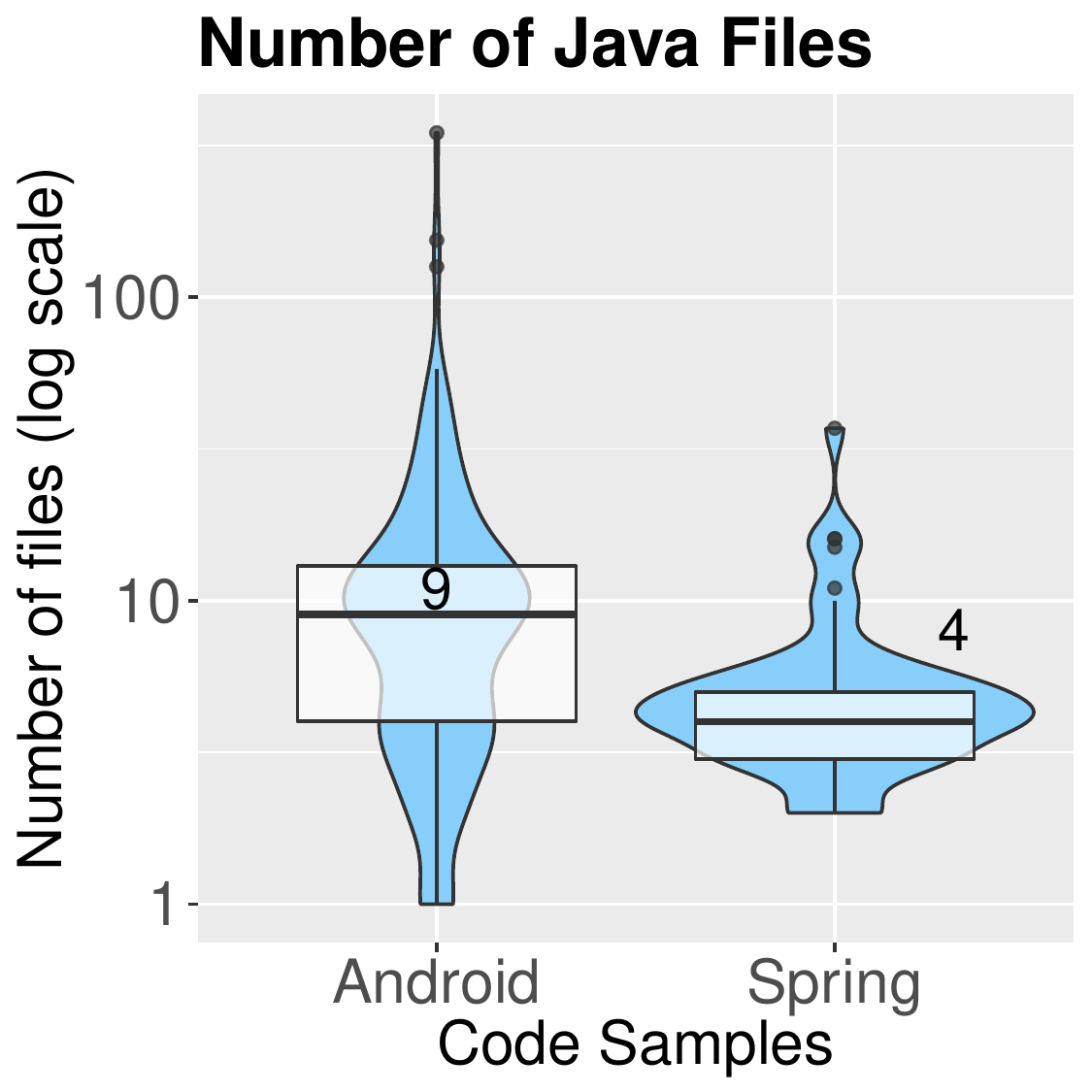}
    \includegraphics[width=0.23\textwidth]{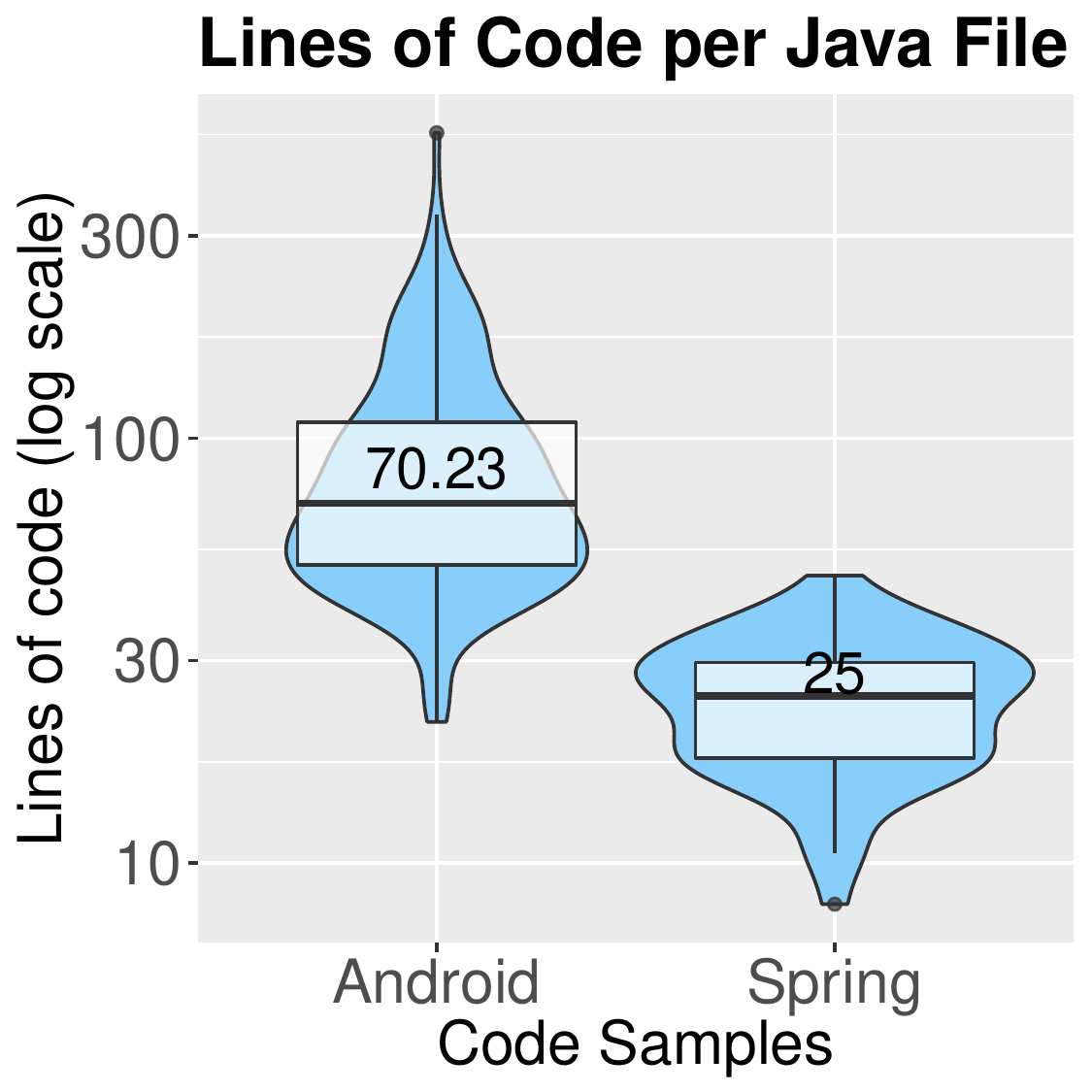}
    \includegraphics[width=0.23\textwidth]{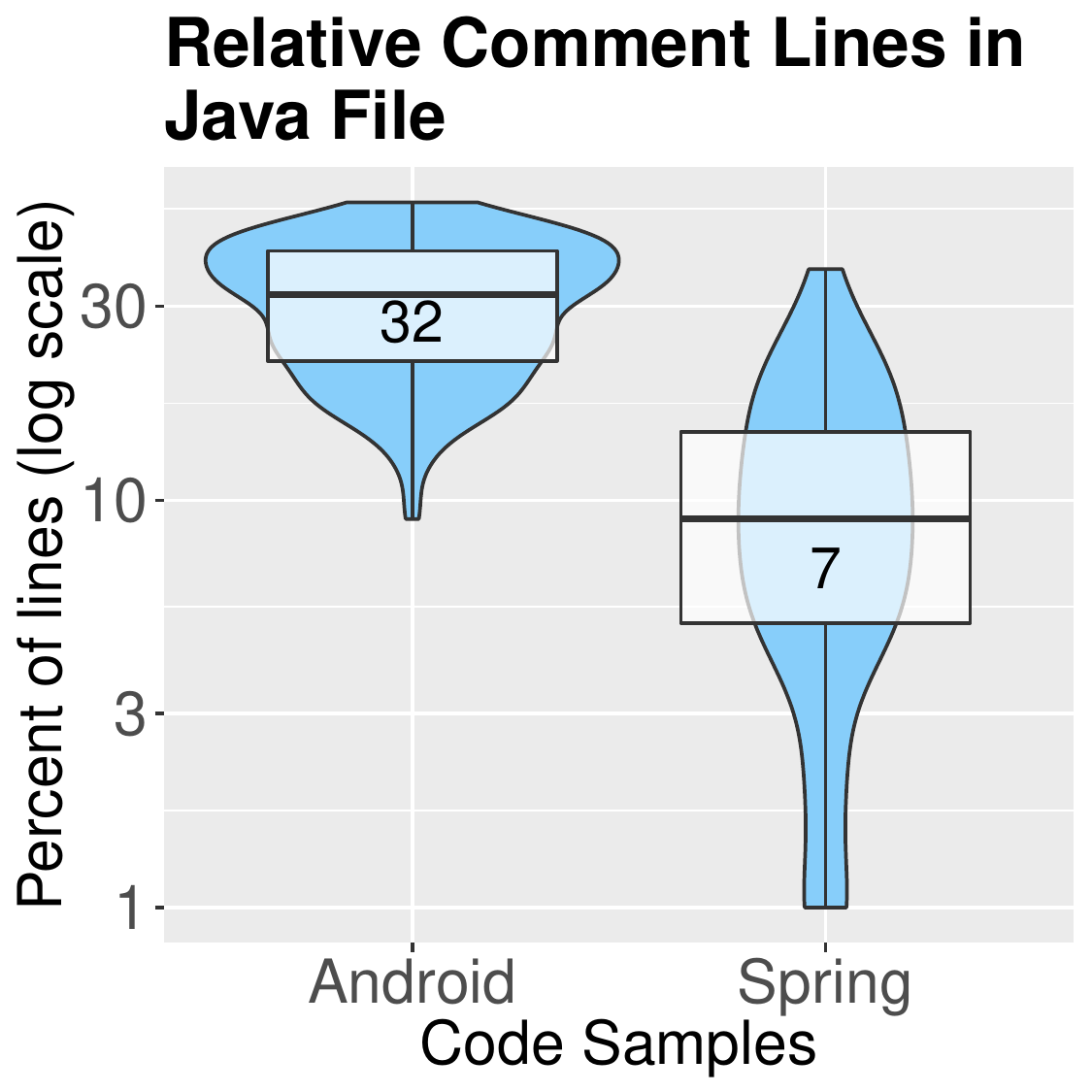}
    \includegraphics[width=0.23\textwidth]{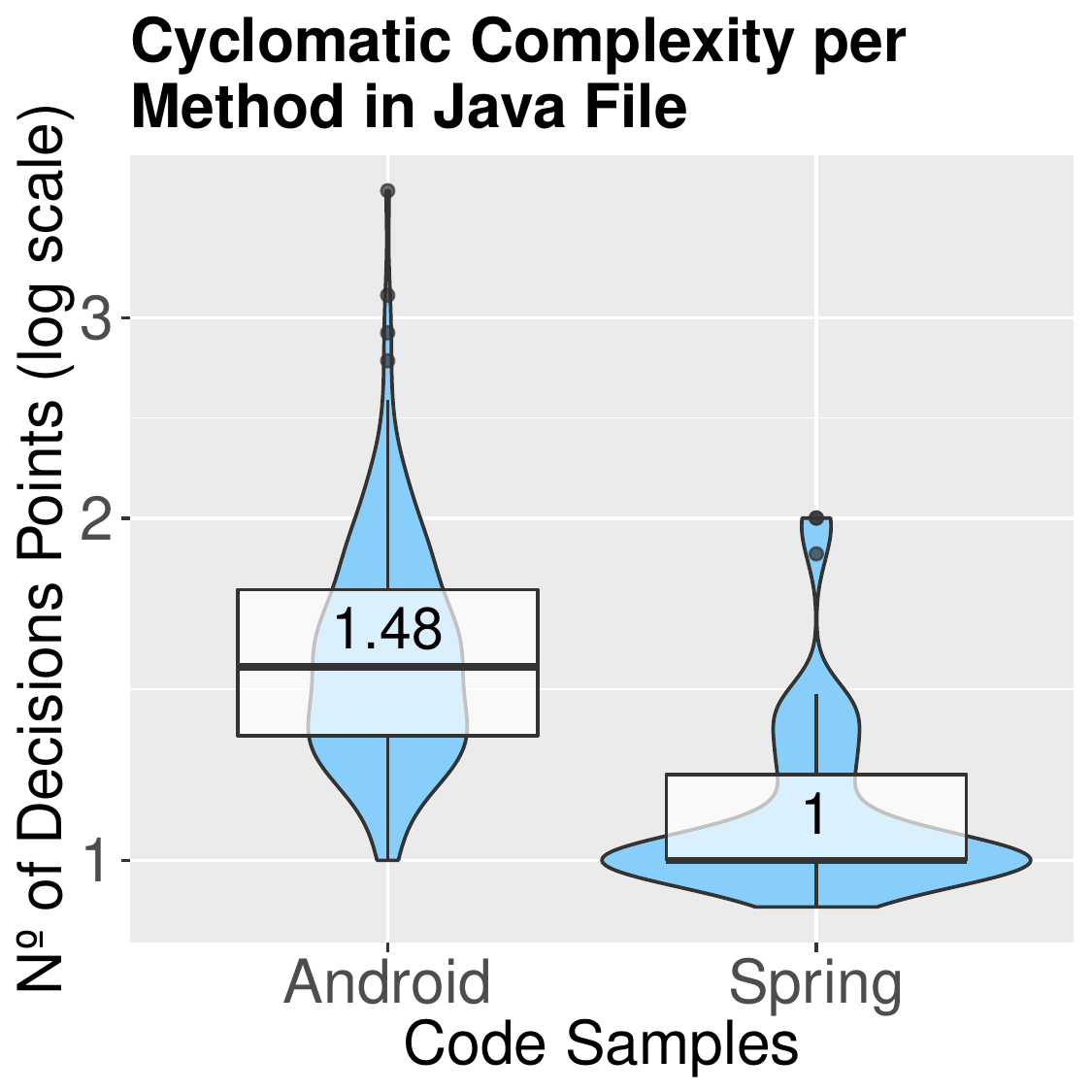}
    \caption{Source code metrics (RQ1).}
    \label{fig:rq1-code-metrics}
\end{figure}

\noindent\emph{Source code metrics:}
Figure~\ref{fig:rq1-code-metrics} presents the distribution of the source code metrics in number of java files, lines of code, cyclomatic complexity, and commented code lines in the last version of the code samples.
We notice that in terms of java files, the projects are very small: on the median 9 files in the Android samples and only 4 in the SpringBoot samples.
The number of lines of code per Java file is larger in Android (70.23) than in SpringBoot (25).
However, the Android samples have more comments (32\%) per file than the SpringBoot samples (7\%).
Finally, we see that the complexity is a bit higher in Android than in SpringBoot samples (1.48 vs. 1).
These numbers confirm our initial impression that code samples are overall small and simple, as stated by guidelines.
However, we also detect that the Android samples are larger and slightly more complex than the SpringBoot ones.

\smallskip

\noindent\emph{File extensions:}
Table~\ref{tab:rq1-file-ext} presents the file extensions found in the analyzed samples.
The Android samples are dominated by xml (15\%), followed by java (9.05\%) and jar files (3.96\%).
The SpringBoot samples include mostly Java (12.49\%), properties (9.75\%), and jar files (8.65\%).
Interestingly, in addition to the java files, both samples provide a relevant proportion of xml and jar files, indicating that a working environment is also available to the clients.

\begin{table}[h]
        \centering
        \footnotesize
        \caption{File extensions (RQ1).}
        \begin{tabular}{lcc | lcc}
        \rowcolor{gray!15}
            \multicolumn{3}{c}{\textbf{Android}} & \multicolumn{3}{c}{\textbf{Spring}} \\ \toprule
            \rowcolor{gray!15}
            Extensions  & \# & \% & Extensions &
            \# & \% \\
            \midrule
                xml        & 4,307  & 15.73  & java        & 319  & 12.49 \\
                java       & 2,477  & 9.05  & properties       & 249   & 9.75  \\
                jar        & 1,083  & 3,96  & jar & 221   & 8.65  \\
                md       & 572  & 2.09  & xml        & 147   & 5.75  \\
                json & 549  & 2,00  & adoc        & 122   & 4.77  \\
                other      & 17,245 & 62.98 & other      & 915 & 35.81
        \end{tabular}
        \label{tab:rq1-file-ext}
    \end{table}

\smallskip

\noindent\emph{Configuration files:}
Table~\ref{tab:rq1-config-files} complements the previous analysis by showing specific configuration files.
Both projects have \texttt{build.gradle} files, which automate software build and delivery via the Gradle Build Tool.
In addition, the SpringBoot samples contains \texttt{pom.xml} files, which relies on Maven and provide features equivalent to Gradle to automate the build process.
The Android samples include the \texttt{manifest.xml} files, which are mandatory to Android apps and provide information that a device needs to run the app.
Finally, to provide continuous integration via the Travis CI,  SpringBoot samples include \texttt{travis.yml} files.

Overall, we notice that both samples include configuration files to support their clients as well as adopt automation tools to improve overall quality~\cite{DuvallMatyasGlover07, Meyer2014, Vasilescu2015}.

\begin{table}[h]
        \centering
        \footnotesize
        \caption{Configuration files (RQ1).}
        \begin{tabular}{lcc | lcc}
        \rowcolor{gray!15}
            \multicolumn{3}{c}{\textbf{Android}} & \multicolumn{3}{c}{\textbf{Spring}} \\ \toprule
            \rowcolor{gray!15}
            Files  & \# & \% & Files &
            \# & \% \\
            \midrule
                build.gradle & 604 & 2.21 & pom.xml      & 144 & 5.64 \\
                manifest.xml       & 397 & 1.45 & build.gradle & 118 & 4.62 \\
                travis.yml   & 2   & 0.01 & travis.yml         & 56 & 2.19 \\
        \end{tabular}
        \label{tab:rq1-config-files}
    \end{table}
    
\begin{tcolorbox}[left=0mm,right=0mm,boxrule=0.1mm,colback=gray!30!white]
\vspace{-0.1cm}
\emph{Lesson Learned 1:} Framework code samples are overall simple and small.
We also find that code samples rely on tools to automate build and integration (\eg~Gradle, Maven, and Travis) and provide a working environment to the users (\ie~including jar, xml, properties, and other files in addition to source code).
\vspace{-0.1cm}
\end{tcolorbox}

\subsection{Evolution (RQ2)}

\begin{figure}[ht]
    \centering
    \includegraphics[width=0.24\textwidth]{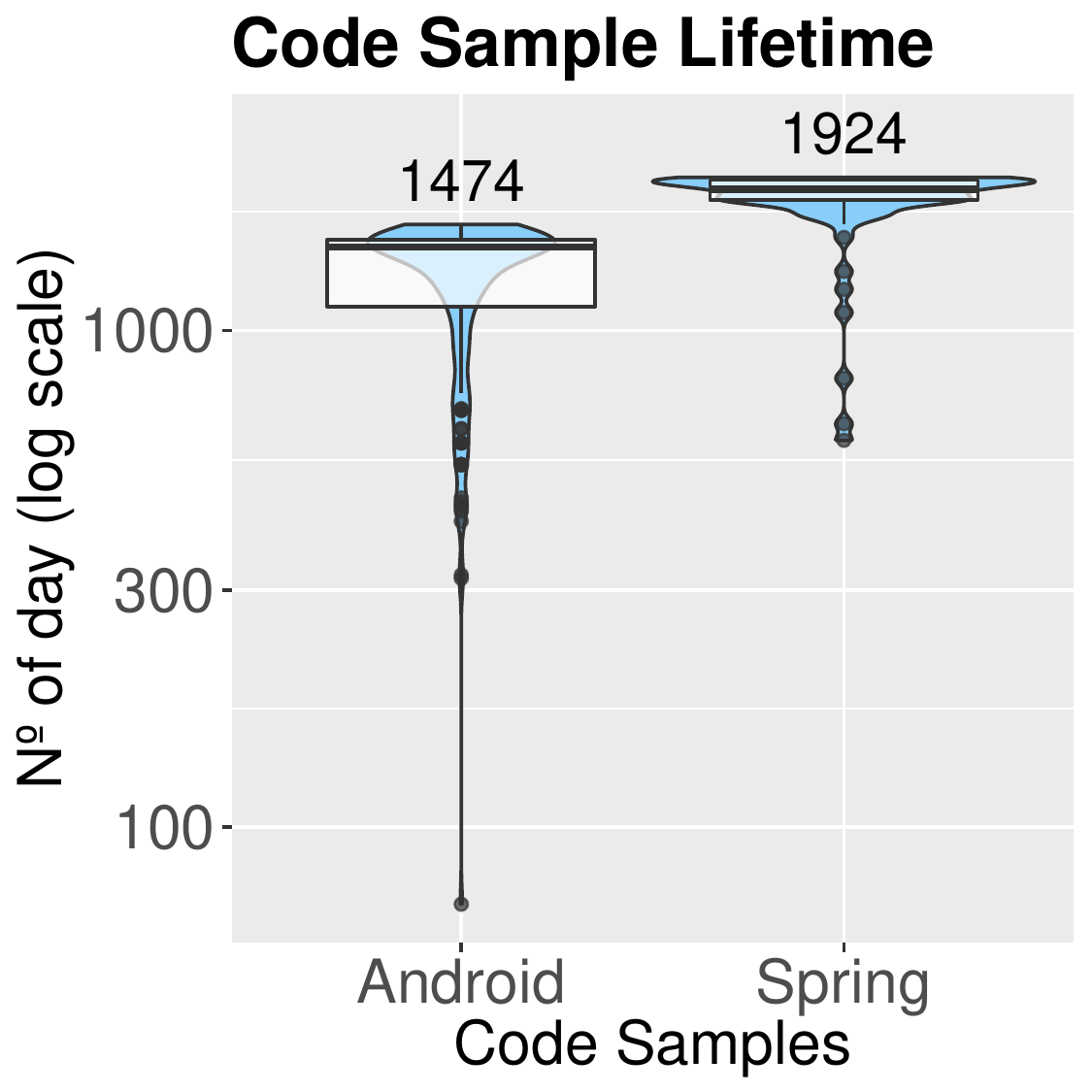}
    \includegraphics[width=0.24\textwidth]{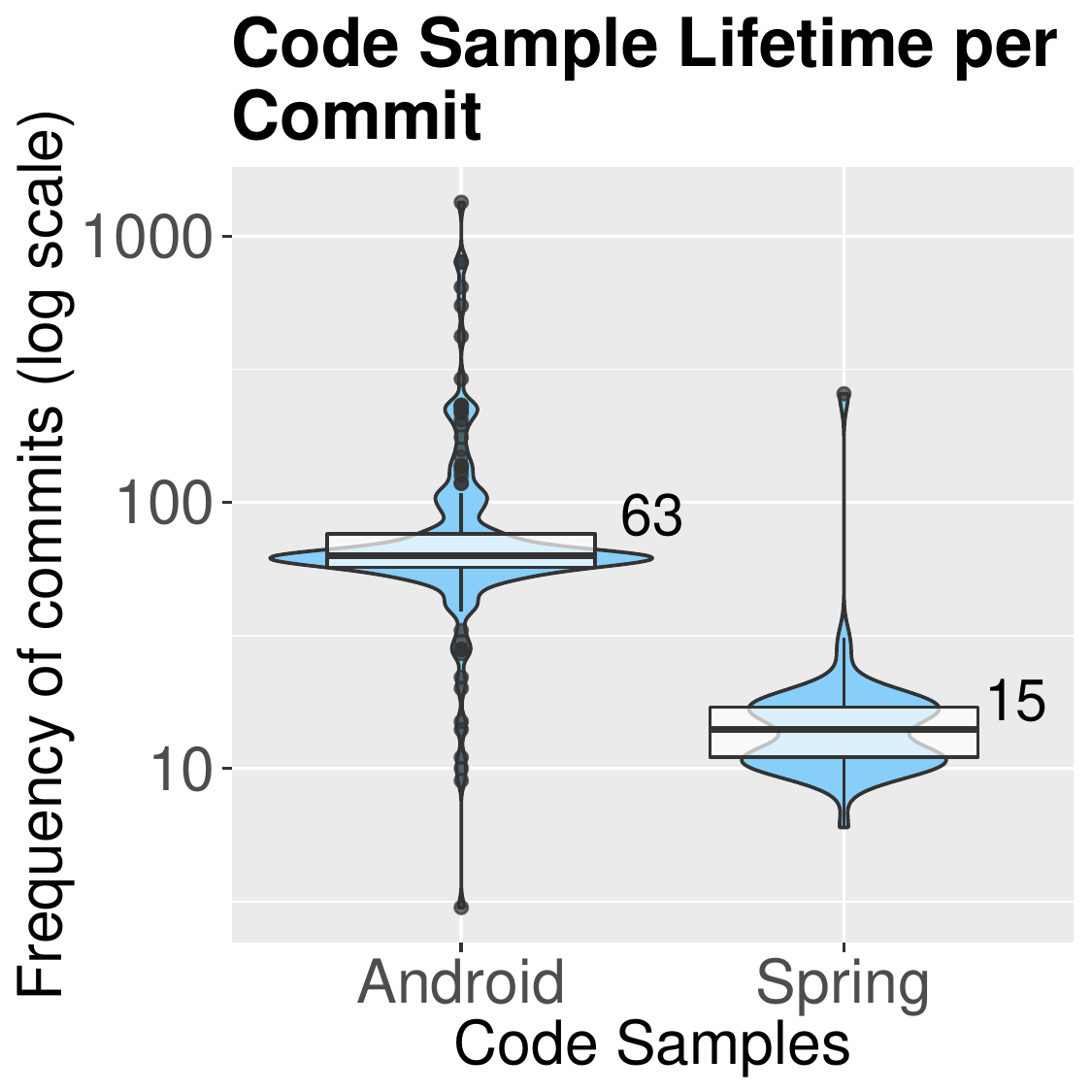}
    \caption{Evolutionary metrics (RQ2).}
    \label{fig:rq2:evol-metrics}
\end{figure}

\noindent\emph{Evolutionary metrics:}
Figure~\ref{fig:rq2:evol-metrics} presents the evolutionary metrics extracted from our samples: lifetime and frequency of commits.
Differently from the previous analysis, \ie~RQ1, these metrics are computed taking into account the code sample changes over time.
We notice that both samples are relatively aged: on the median, the Android samples have 1,474 days (4 years), while the SpringBoot ones are even older, having 1,924 days (5.2 years).
Regarding the frequency of commits, the Android samples change one time each 63 days, while the SpringBoot one time each 15 days, on the median.

\smallskip

\noindent\emph{File extension changes:}
Table~\ref{tab:rq2-file-evol} presents the changes per file extension.
We clearly see that the code samples are not static: they are updated over the years.
In both cases, xml files are the most changed, followed by Java, properties, and jar files.
Table~\ref{tab:rq2-change-type} shows another view of this data: the  actions performed on the files: addition, modification, or removal.
While in Android samples most of the actions are to add files (53.03\%), in SpringBoot the majority is to modify existing ones (85.13\%).
In both cases, removal of files is uncommon.

\begin{table}[h]
        \centering
        \footnotesize
        \caption{File extension changes (RQ2).}
        \begin{tabular}{lcc | lcc}
        \rowcolor{gray!15}
            \multicolumn{3}{c}{\textbf{Android}} & \multicolumn{3}{c}{\textbf{Spring}} \\ \toprule
            \rowcolor{gray!15}
            Extensions  & \# & \% & Extensions &
            \# & \% \\
            \midrule
                xml        & 9,075  & 15.67 & xml        & 7,735  & 28.75 \\
                java       & 7,034  & 12.14 & java       & 1,437  & 5.34  \\
                properties & 1,926  & 3.33  & properties & 961   & 3.57  \\
                jar        & 1,783  & 3.08  & jar        & 770   & 2.86  \\
                json       & 1,111  & 1.92  & bat        & 331   & 1.23  \\
                other      & 36,988 & 63.86 & other      & 15,666 & 58.24
        \end{tabular}
        \label{tab:rq2-file-evol}
    \end{table}
    
    \begin{table}[]
        \centering
        \footnotesize
        \caption{Action type per file (RQ2).}
        \begin{tabular}{lcc | lcc}
        \rowcolor{gray!15}
            \multicolumn{3}{c}{\textbf{Android}} & \multicolumn{3}{c}{\textbf{Spring}} \\ \toprule
            \rowcolor{gray!15}
            File action type  & \# & \% & File action &
            \# & \% \\
            \midrule
                Add    & 30,716 & 53.03  & Modify & 22,900 & 85.13  \\
                Modify & 23,696 & 40.91  & Add    & 3,020  & 11.23  \\
                Delete & 3,505  & 6.05   & Delete & 980   & 3.64   \\
                \midrule
                Total  & 57,917 & 100.00 & Total  & 26,900 & 100.00
        \end{tabular}
        \label{tab:rq2-change-type}
\end{table}

\smallskip
    
\noindent\emph{Configuration file changes:}
Table~\ref{tab:rq2-conf-evol} presents the most changed configuration files.
We notice that \texttt{build.gradle} files are the most changed in both frameworks.
In Android code samples, the \texttt{manifest.xml} are usually changed, while in SpringBoot the \texttt{pom.xml} are often updated.
Therefore, as most of these files are related to automation tools, we can confirm that these tools keep being updated over time.

\begin{table}[h]
        \centering
        \footnotesize
        \caption{Configuration file changes (RQ2).}
        \begin{tabular}{lcc | lcc}
        \rowcolor{gray!15}
            \multicolumn{3}{c}{\textbf{Android}} & \multicolumn{3}{c}{\textbf{Spring}} \\ \toprule
            \rowcolor{gray!15}
            Files  & \# & \% & Files &
            \# & \% \\
            \midrule
                build.gradle & 5,281 & 9.12 & build.gradle & 7,565 & 28.12 \\
                manifest.xml & 1,076 & 1.86 & pom.xml      & 7,531 & 28.00 \\
                travis.yml   & 24   & 0.04 & travis.yml   & 208  & 0.77  \\
        \end{tabular}
        \label{tab:rq2-conf-evol}
\end{table}

\smallskip 

\begin{figure}[h]
        \centering
        \includegraphics[width=0.31\textwidth]{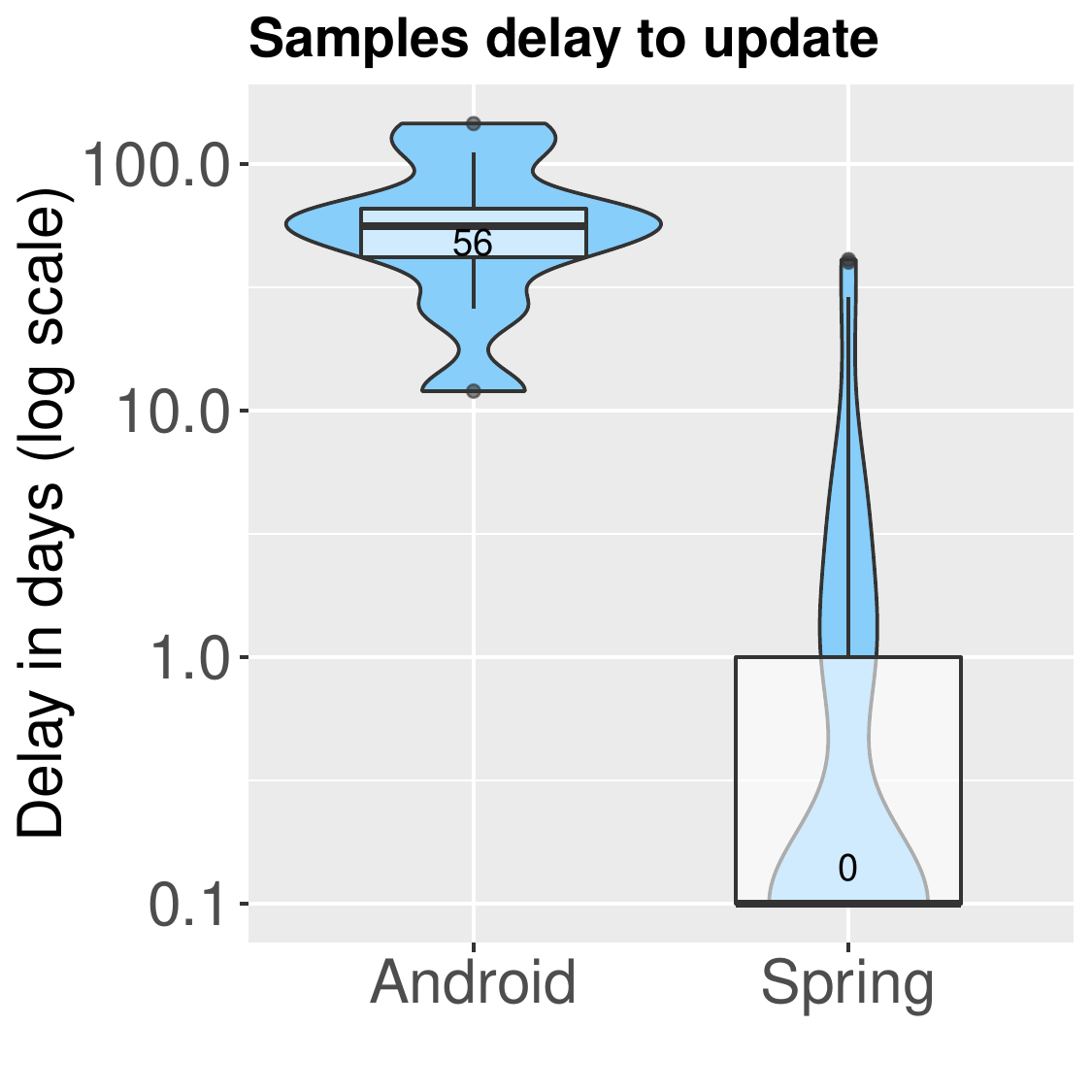}
        \caption{Migration delay (RQ2)}
        \label{fig:rq2-delay}
\end{figure}

\noindent\emph{Migration delay:}
Figure~\ref{fig:rq2-delay} presents the delay in number of days the sample take to migrate to new versions of the Android and SpringBoot frameworks.
SpringBoot samples migrate much quicker than Android ones.
While SpringBoot samples update in same day the new version is available (median zero days), the Android samples take 56 days to migrate, on the median.

\begin{figure}[h]
    \centering
    \includegraphics[width=0.24\textwidth]{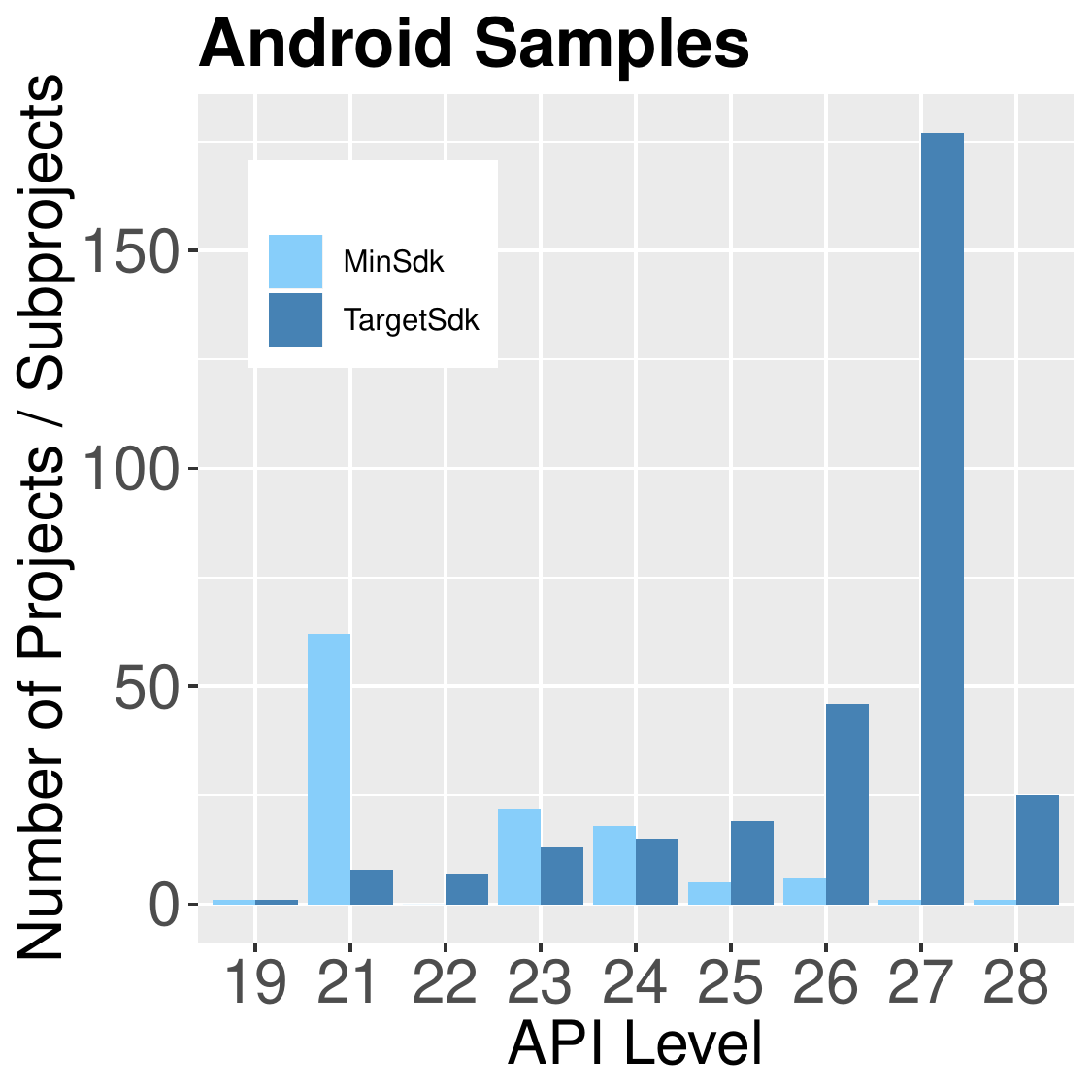}
    \includegraphics[width=0.24\textwidth]{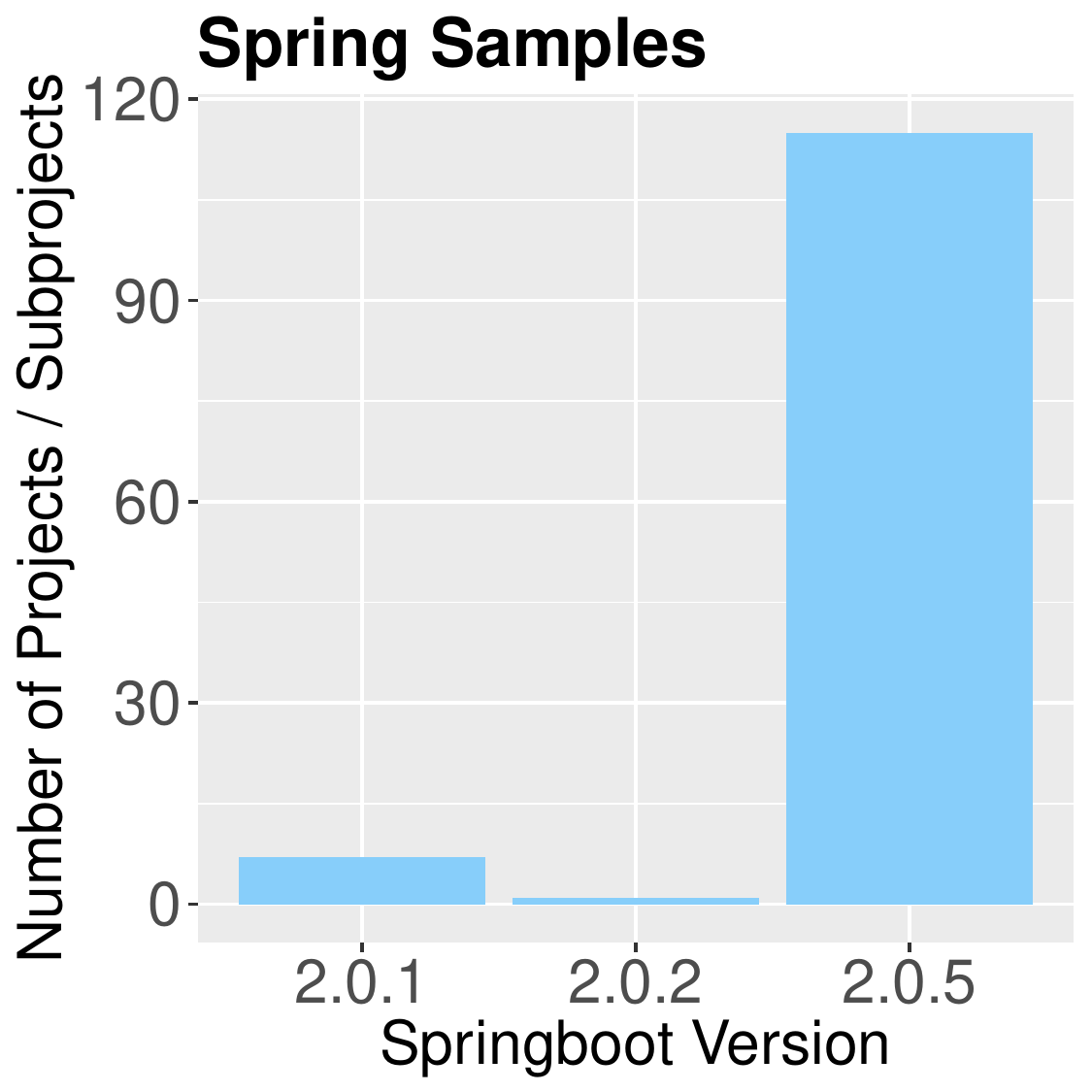}
    \caption{Code sample versions (RQ2).}
    \label{fig:rq2-versions}
\end{figure}

Figure~\ref{fig:rq2-versions} shows the versions that the code samples are adopting.
We see that the Android samples mostly rely on\footnote{That is, they have the \texttt{TargetSdk} set to a certain version.} the API level 26 (\ie~Android 8.0, Oreo), 27 (\ie~8.1, Oreo), and 28 (\ie~9.0, Pie), however, many samples also rely on other API levels, which represents older versions of Android.
Regarding SpringBoot, the majority of the samples are based on version 2.0.5; in this case, we found no sample relying on versions under 2.0, which represents older SpringBoot versions.

\begin{figure}[h]
\centering
    \includegraphics[width=0.24\textwidth]{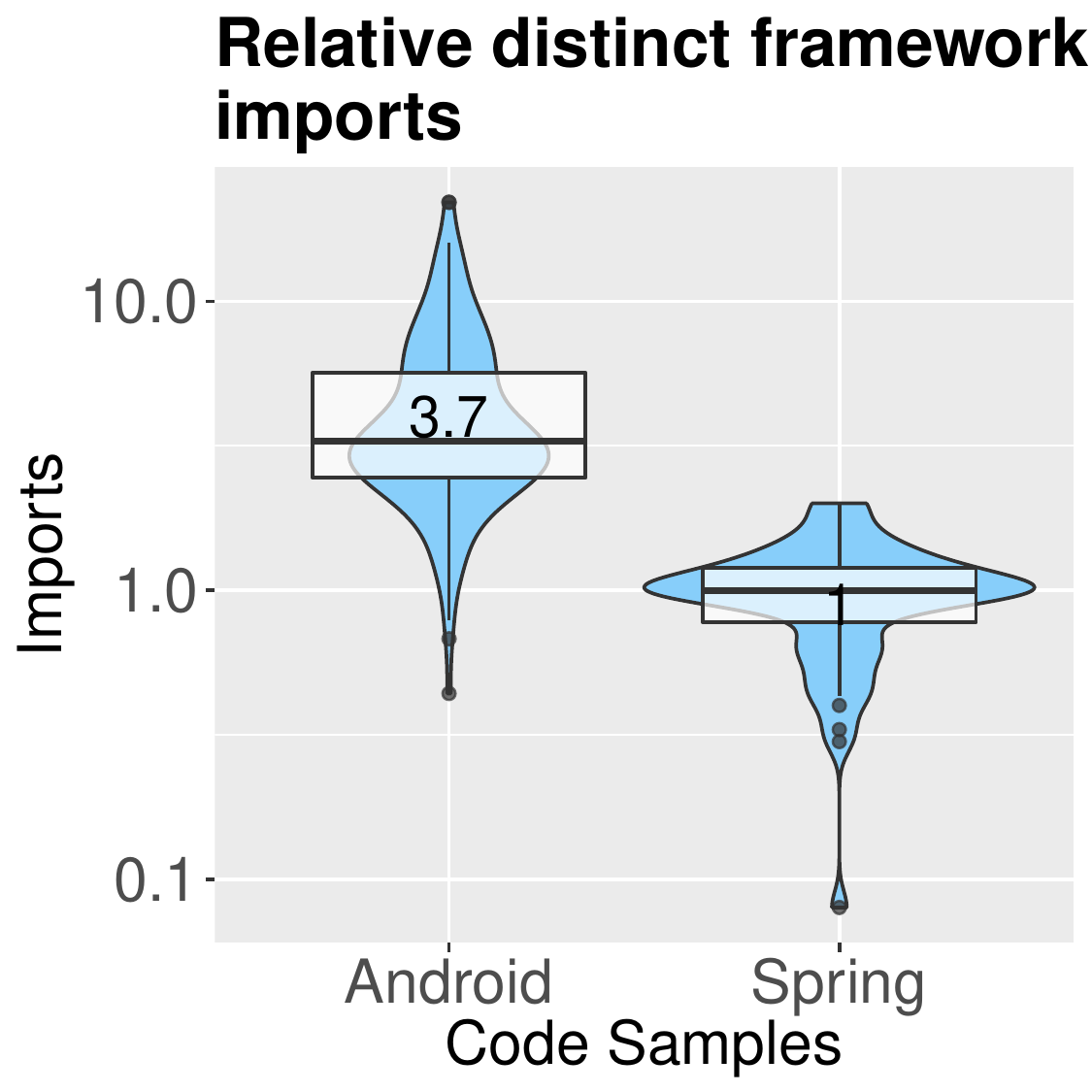}
    \includegraphics[width=0.24\textwidth]{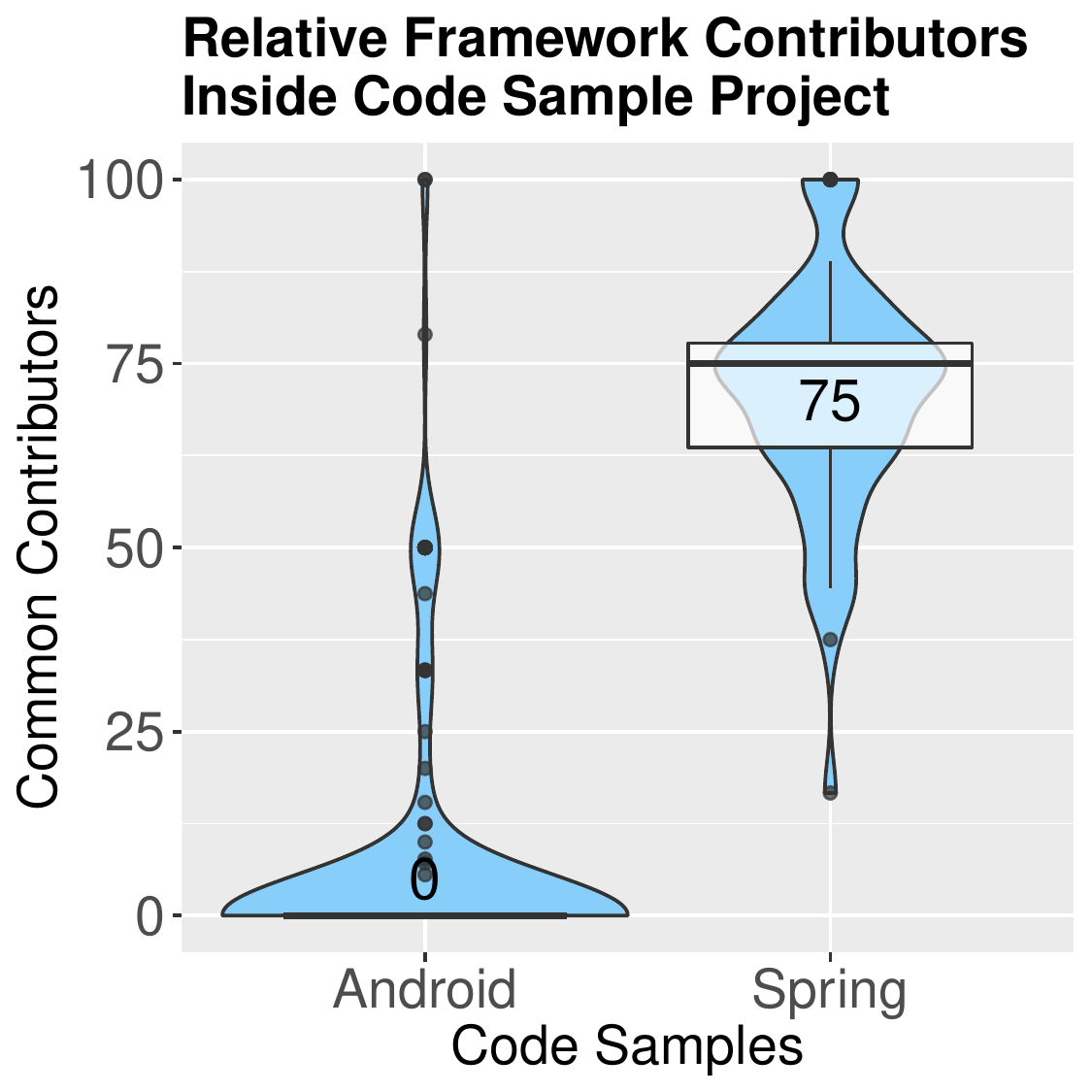}
\caption{Dependency to the framework in number of imports (left) and ratio of developers in both code samples and framework (right).}
\label{fig:rq2-etc}
\end{figure}

To better understand the reason the SpringBoot code samples are migrated faster than the Android ones, we investigated two scenarios.
First, we hypothesize that the Android code samples are more complex than SpringBoot ones.
Indeed, we have seen in RQ1 that Android code samples are slightly more complex.
In addition, Figure~\ref{fig:rq2-etc} (left) presents another view of complexity and shows that the Android code samples rely more on Android APIs than the SpringBoot ones (3.7 vs. 1, on the median).
Thus, it is natural that migration takes longer in Android code samples as they are more coupled to the framework.
Our second hypothesis is that the developers who maintain the SpringBoot code samples are the same who maintain the SpringBoot framework itself.
Figure~\ref{fig:rq2-etc} (right) shows the ratio of developers working on both code samples and framework.
We notice that ratio is quite large in SpringBoot: on the median, 75\% of the developers who commit code in the samples have also committed in the framework SpringBoot; in Android, this ratio is zero.
Therefore, having developers working on both code samples and framework may support their maintenance by decreasing migration delay.

\smallskip

\begin{tcolorbox}[left=0mm,right=0mm,boxrule=0.1mm,colback=gray!30!white]
\vspace{-0.1cm}
\emph{Lesson Learned 2:} Code samples are not static, but they evolve over time. 
Updates are made on both source code and configuration files, for example, to keep them up to date with new framework versions.
Overall, code samples are migrated quickly and often rely on recent framework versions.
Moreover, having developers working on both code samples and framework may decrease the migration delay.
\vspace{-0.1cm}
\end{tcolorbox}

\subsection{Popularity (RQ3)}

Table~\ref{tab:rq3-pop} presents the results for the popularity analysis.
The popular and unpopular Android code samples are statistically significant different regarding the metrics \emph{java files}, \emph{lines of code}, and \emph{cyclomatic complexity}, all with medium effect.
The metric \emph{frequency of commits} is also distinct, but with small effect, that is, popular Android samples have statistically significant more changes in shorter periods than the unpopular ones.
In SpringBoot, we do not find any difference among the popular and unpopular code samples with respect to the investigated metrics.

\begin{table}[h]
        \centering
        \footnotesize
        \caption{Popularity analysis (RQ3). Comparison between popular and unpopular samples (Pop x Unp). Statistically significant difference with small (S) or medium (M) effect. Direction of the difference (Dir)}
        
        \begin{tabular}{l cc | cc}
            \rowcolor{gray!15}
            & \multicolumn{2}{c|}{\textbf{Android}} & \multicolumn{2}{c}{\textbf{Spring}} \\ \toprule
            \rowcolor{gray!15}
            Metrics & Pop x Unp & Dir & Pop x Unp & Dir \\ 
            \midrule
            Java files & $\leq$0.001 (M) & $\uparrow$ & 0.15 & -\\
            Lines of Code & $\leq$0.001 (M) & $\uparrow$ & 0.60 & -\\
            Relative comment lines & 0.57 & - & 0.42 & - \\
            Cyclomatic Complexity & $\leq$0.001 (M) & $\uparrow$ & 0.42 & -\\
            \midrule
            Lifetime & 0.38 & - & 0.28 & - \\
            Frequency of commits & $\leq$0.001 (S) &$\downarrow$ & 0.08 & - \\
        
            \bottomrule
        \end{tabular}
        \label{tab:rq3-pop}
\end{table}

\begin{tcolorbox}[left=0mm,right=0mm,boxrule=0.1mm,colback=gray!30!white]
\vspace{-0.1cm}
\emph{Lesson Learned 3:} 
Popular Android code samples have a higher amount of code files, are longer and more complex, and have more changes over time.
\vspace{-0.1cm}
\end{tcolorbox}

\subsection{Client Usage (RQ4)}

\begin{figure}[h]
        \centering
        \includegraphics[width=0.23\textwidth]{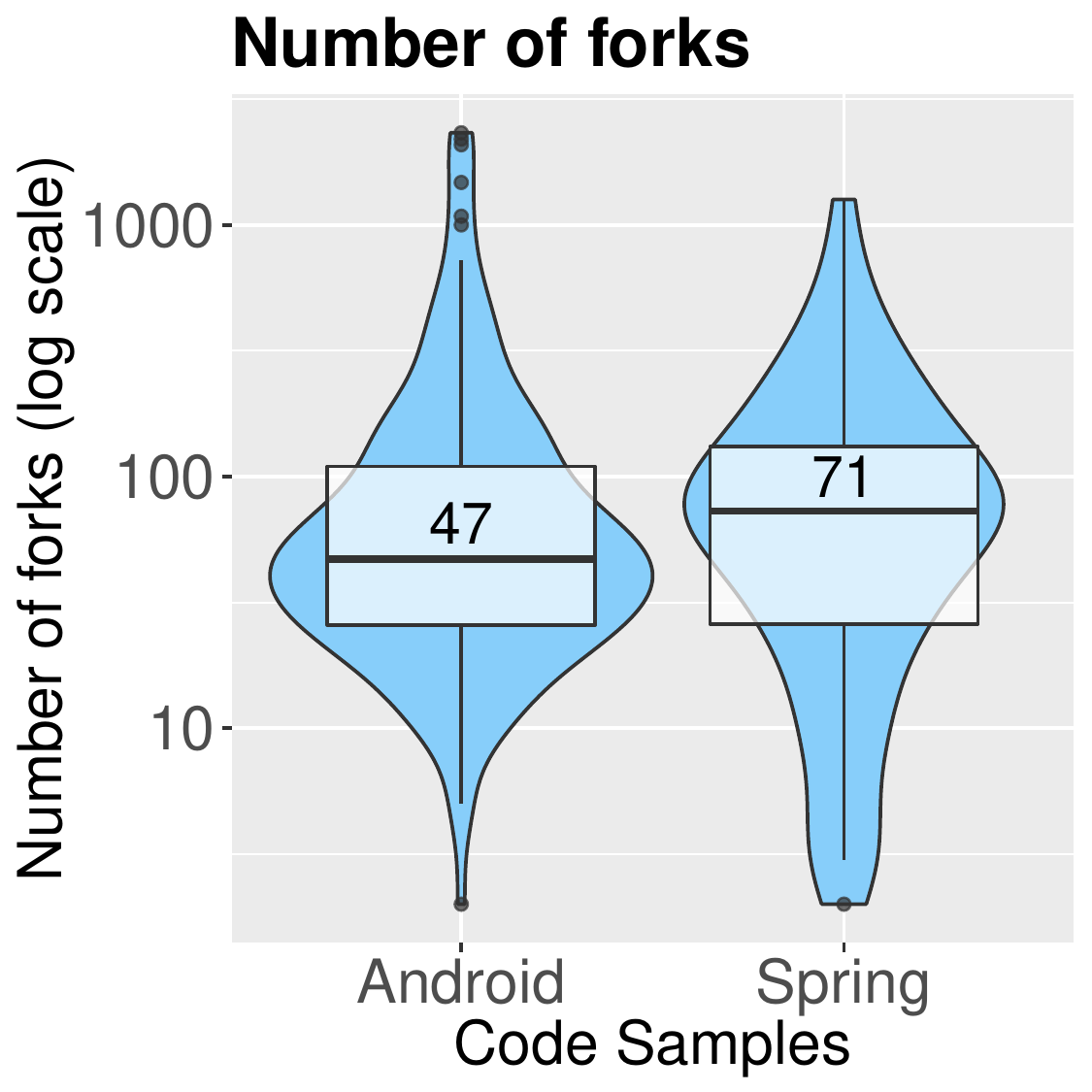}
        \includegraphics[width=0.23\textwidth]{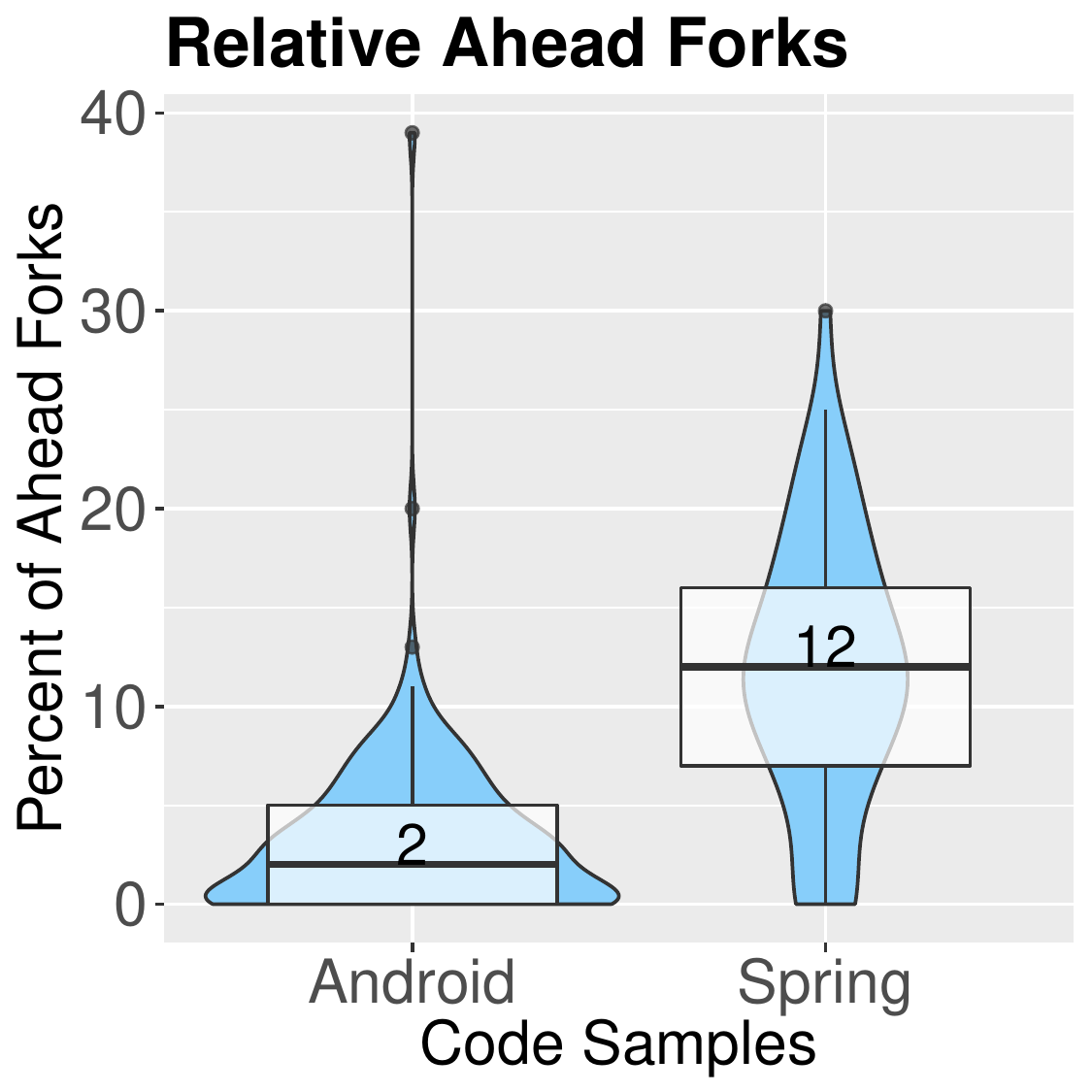}
        \caption{Code sample forks (RQ4).}
        \label{fig:rq4-forks}
\end{figure}

\noindent\emph{Fork metrics:}
We adopt the fork metric as a proxy of client usage for the code samples.
We detected 25,106 forks of Android code samples and 7,025 of SpringBoot ones.
Figure~\ref{fig:rq4-forks} (left) presents the distribution of the number forks per code sample.
We see that Android code samples have on the median 47 forks, while the third quartile is 112.
In SpringBoot code samples, the median is 71 forks and the third quartile is 137.5.
The most forked code sample in Android is \texttt{android-testing} (2,409 forks), while in SpringBoot the most forked is \texttt{gs-rest-service} (1,412  forks).

The fact that there is a fork do not necessarily mean that it changes over time.
Indeed, in Android, only 3\% (871 out of 25,106) forked projects are ahead of the base project, \ie~they performed at least one commit; in SpringBoot this ratio is 15\% (1,055 out of 7,025).
Figure~\ref{fig:rq4-forks} (right) presents the distribution forked code samples with commits.
On the median, only 2\% of the forked Android code samples have commits; in SpringBoot, this ratio is higher: 12\%.
Overall, we notice that most of the forked code samples are inactive.

\begin{figure}[h]
    \centering
    \includegraphics[width=0.33\textwidth]{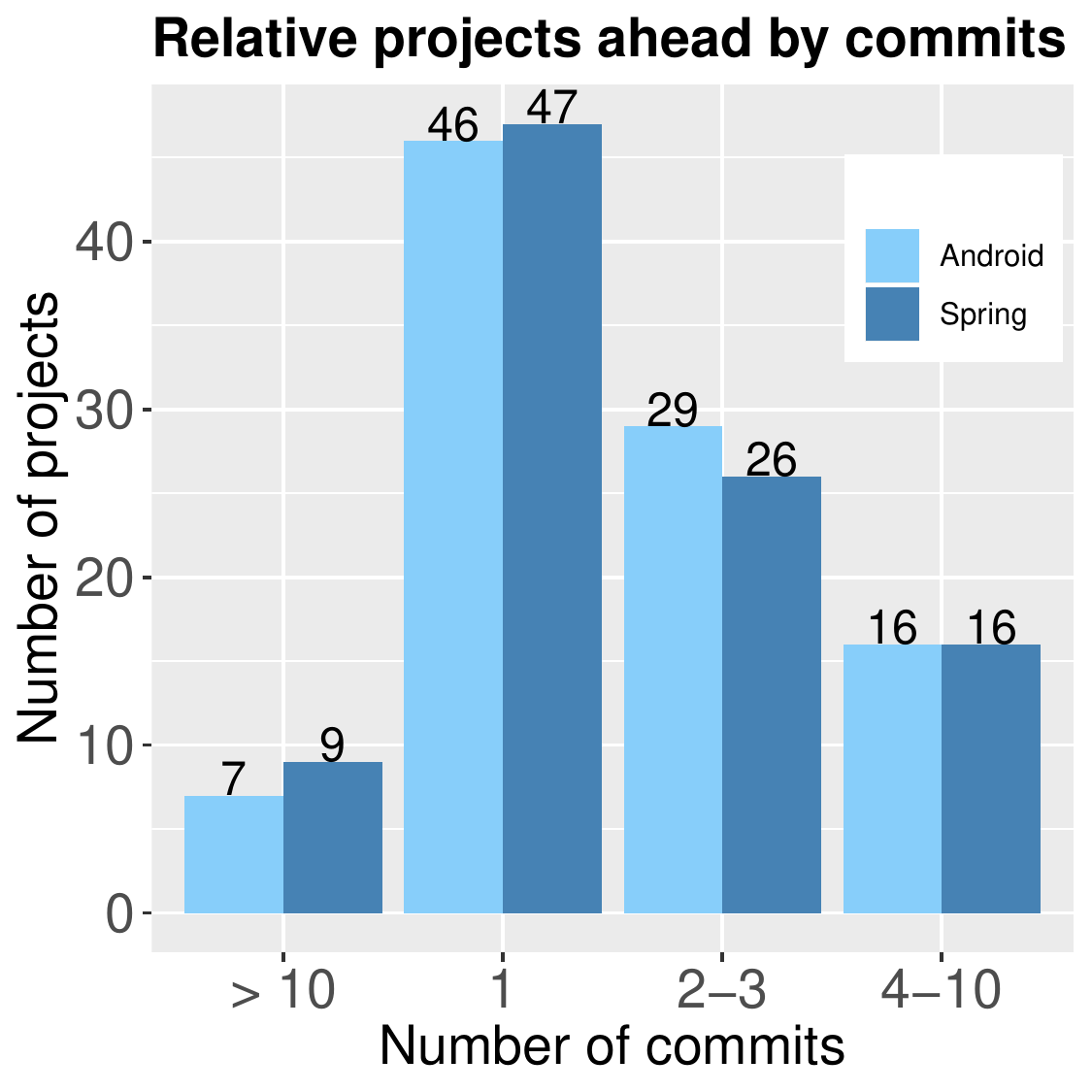}
    \caption{Commits in forked code samples (RQ4).}
    \label{fig:rq4:ahead-by-commit}
    \end{figure}

Figure~\ref{fig:rq4:ahead-by-commit} presents the frequency of commits per forked code samples; here, we only show the forks with at least one commit.
In this case, 7\% and 9\% of the forked Android and SpringBoot code samples have 10 or more commits.
In both frameworks, the majority of the forked code samples have a single commit (46\% and 47\%).
In Android, 29\% of the forked code samples have 2--3 commits, while 16\% have 4--10.
In SpringBoot, the ratios are equivalent: 26\% have 2--3 commits while 16\% have 4--10.

\smallskip
    
\noindent\emph{File extension changes in forked code samples:}
Table~\ref{tab:rq4:file-ext} shows the file extension changes in the forked code samples.
The notice that the developers change mostly xml, json, java, and jar files.
Table~\ref{tab:rq4-actions} shows the actions performed on the files: addition, modification, or removal.
While in Android samples most of the actions are to add files (56.97\%), in SpringBoot the majority is to modify existing ones (43.59\%).

\begin{table}[h]
        \centering
        \footnotesize
        \caption{File extension changes in forked code samples (RQ4).}
        \begin{tabular}{lcc | lcc}
        \rowcolor{gray!15}
            \multicolumn{3}{c}{\textbf{Android}} & \multicolumn{3}{c}{\textbf{Spring}} \\ \toprule
            \rowcolor{gray!15}
            Extensions  & \# & \% & Extensions &
            \# & \% \\
            \midrule
                xml   & 24,022 & 17.39 & java       & 4,525 & 34.56 \\
                json  & 8,530  & 6.17  & xml        & 1,128 & 8.61  \\
                java  & 8,298  & 6.01  & jar        & 983  & 7.51  \\
                jar   & 3,784  & 2.74  & properties & 709  & 5.41  \\
                txt   & 1,264  & 0.91  & yml        & 398  & 3.04  \\
                other & 92,253 & 66.78 & other      & 5,352 & 40.87
        \end{tabular}
        \label{tab:rq4:file-ext}
    \end{table}

\begin{table}[h]
        \centering
        \footnotesize
        \caption{Action type per file in the forked code samples (RQ4).}
        \begin{tabular}{lcc | lcc}
        \rowcolor{gray!15}
            \multicolumn{3}{c}{\textbf{Android}} & \multicolumn{3}{c}{\textbf{Spring}} \\ \toprule
            \rowcolor{gray!15}
            File Action  & \# & \% & File Action &
            \# & \% \\
            \midrule
                Add    & 78,706  & 56.97  & Modify & 5,708  & 43.59  \\
                Delete & 42,971  & 31.10  & Add    & 4,640  & 35.43  \\
                Modify & 16,474  & 11.92  & Delete & 2,747  & 20.98  \\
                \midrule
                Total  & 138,151 & 100.00 & Total  & 13,095 & 100.00
        \end{tabular}
        \label{tab:rq4-actions}
    \end{table}
    
\begin{tcolorbox}[left=0mm,right=0mm,boxrule=0.1mm,colback=gray!30!white]
\vspace{-0.1cm}
\emph{Lesson Learned 4:} The majority of the forked code samples are inactive.
However, a non-negligible percentage of the forked code samples are updated and evolve over time.
The changes are mostly concentrated in xml and java files.
\vspace{-0.1cm}
\end{tcolorbox}

\section{Implications}
\label{sec:summary}

Based on our findings, we provide a set of implications to framework code sample creators and clients in order to support their maintenance and usage practices:

\smallskip

\checkmark Code samples should be simple and small to facilitate their reuse, as stated by good development practices~\cite{mozilla18}.
Indeed, the majority of the code samples provided by Android and SpringBoot follow this rule.
However, this is not strict: we find that the code samples with more java files are more likely to be popular in Android.

\smallskip

\checkmark Code samples should provide working environments to ease their usage.
Indeed, most of the Android and SpringBoot code samples are formed by source code and many other configuration files necessary to properly run them.
Automated build and integration tools may also support both the creators and clients, improving their quality and reducing risks~\cite{DuvallMatyasGlover07, Meyer2014, Vasilescu2015}.

\smallskip

\checkmark Code samples are not frozen projects, but they should be updated over time.
Changes are commonly performed to follow recent framework versions, otherwise the code samples become out of date and less attractive to the clients~\cite{McDo13, Robb12, laerte17, sqj2018, Kula2018}.
Indeed, this practice is often performed by Android and SpringBoot code samples, but much faster in the latter.
We also find that the code samples that are changed frequently are more likely to be popular in Android.

\smallskip

\checkmark Code samples may benefit from scenarios where their developers also contribute to the framework itself. For example, we found that migration delay may decrease in cases in which the overlap of developers is higher between code samples and framework. We recognize, however, that this phenomenon should be more explored in further research.

\smallskip 

\checkmark The majority of the forked code samples are inactive, however, a non-negligible percentage are updated by their clients as a way to explore and learn them. Thus, we recommend this cycle (fork-change-learn) to the clients kick start in a code sample.

\section{Threats to Validity}
\label{sec:threats}

This section discusses the study limitations based on the four categories of validity threats described by Wohlin~et al.~\cite{Wohlin:2012}. Each category has a set of possible threats to the validity of an experiment. We identified these possible threats to our study within each category, which are discussed in the following with the measures we took to reduce each risk.

\smallskip

\noindent\textit{Conclusion validity:}
It concerns the relationship between the treatment and the outcome. In this work, potential threats arise from \textit{violated assumptions of statistical tests}: the statistical tests used to support our conclusions may have been inappropriately chosen. To mitigate this threat wherever possible, we used statistical tests obeying the characteristics of our data. More specifically, we used non-parametric tests, which do not make any assumption on the underlying data distribution regarding variances and types.

\smallskip

\noindent\textit{Internal validity:}
It is the degree to which conclusions can be drawn about the causal effect of independent variables on the dependent variables. One important threat to the internal validity is related to the \textit{ambiguity about the direction of causal influence}: specifically in RQ3, aspects from code samples may be a key to their popularity. On the other hand, the popularity of a code sample may influence code sample aspects measured in our study as code comments and cyclomatic complexity. To ameliorate this threat, we analyze the history of the code samples in order to avoid considering aspects arisen due to the increase in popularity over time.

\smallskip

\noindent\textit{Construct validity:}
It refers to the degree to which inferences can legitimately be made from the operationalizations in your study to the theoretical constructs on which those operationalizations were based. We detected a possible threat related to the  \textit{restricted generalizability across constructs}: Java might present specific source code characteristics when compared to other programming languages and affects RQ1. This risk cannot be avoided since we analyzed only source code implemented in Java. However, we argue that Java is an important programming language and comprises a large number of code samples in GitHub repository.

\smallskip

\noindent\textit{External validity:}
Threats associated with external validity concern the degree to which the findings can be generalised to the wider classes of subjects from which the experimental work has drawn a sample. We identified a risk related to \textit{the interaction between selection and treatment}: the use of code samples provided by two frameworks might present specific aspects when compared to other frameworks. This risk cannot be avoided because our focus are the two frameworks presented in Section~\ref{sec:design}. However, we argue that they are relevant and worldwide adopted frameworks that have millions of end-users. Therefore, we believe the results extracted can be a first step towards the generalization of the results. 

\section{Related Work}
\label{sec:related}

Frameworks are used to support development, providing source code reuse, improving productivity, and decreasing costs~\cite{Mose96, Kons09, raemaekers12}.
Often there is a steep learning curve involved when developers adopt frameworks. Development based on code samples provides the benefits of code reuse, efficient development, and code quality~\cite{Sindhgatta:2006}. Moreover, with the popularity and relevance of the Question and Answer (Q\&A) sites as Stack Overflow, some studies propose approaches and tools to search and/or retrieve source code samples as well as explore properties of those samples.

\smallskip

\noindent\textit{Context-based code samples.} Software engineering tools are bringing sophisticated search power into the development environment by extending the browsing and searching capabilities~\cite{Holmes:2005,Mandelin:2005,Sindhgatta:2006,Poshyvanyk:2006,Sahavechaphan:2006}. For instance, Holmes~and~Murphy~\cite{Holmes:2005} proposed a technique that recommends source code examples from a repository by matching structures of given code. XSnippet~\cite{Sahavechaphan:2006} provides a context-sensitive code assistant framework that provides sample source code snippets for developers. In general, these tools help locate samples of code, demonstrate the use of frameworks and fasten development by exploring the syntactic context provided mainly by the IDE to recommend code samples more relevant to developers (as in Strathcona~\cite{Holmes:2005}). However, the samples provided by these systems are highly dependent of a particular development context, whereas code samples typically are complete projects that were made to facilitate and accelerate the learning process of features provided by frameworks. Therefore, it is expected that the types of code samples explored in this paper present different characteristics when compared to samples automatically generated by tools.

\smallskip

\noindent\textit{Mining API usage examples.} Complementing the tools aforementioned, many studies confirmed the the significance of API usage examples, mainly in the context of framework APIs, and proposed approaches to mine API usage examples from open code repositories and search engines~\cite{Zhu:2014,Montandon:2013,Moreno:2015,Buse:2012,Keivanloo:2014,Niu:2017}. Most of these work retrieve the so-called code snippets to support API learning, whereas our work focus on complete projects of framework code samples. In addition, our work is not focused on proposing an approach to mine code samples, but analyze characteristics of these code samples.

\smallskip

\noindent\textit{Assessing Q\&A code snippets.} Nasehi~et~al.~\cite{Nasehi:2012} focused on finding the characteristics of a good example on Stack Overflow. They adopted an approach based on high/low voted answers, number of code blocks used, the conciseness of the code, the presence of links to other resources, the presence of alternate solutions, and code comments. Yang~et~al.~\cite{Yang:2016} assessed the usability of code snippets across four languages: C\#, Java, JavaScript, and Python. The analysis was based on the standard steps of parsing, compiling and running the source code, which indicates the effort that would be required for developers to use the snippet as-is. Finally, there are studies analyzing the adoption of code snippets~\cite{Roy:2010,Heinemann:2011,Yang:2017}. For instance, Roy~and~Cordy~\cite{Roy:2010} analyzed code snippet clones in open source systems. They found that on average 15\% of the files in the C systems, 46\% of the files in the Java systems and 29\% of files in the C\# systems are associated with exact (block-level) clones. Similar to our work, these studies focus on analyzing properties of code snippets and their adoption on real projects. However, our work targets entire code sample projects instead of code snippets.

\section{Conclusion}
\label{sec:conclu}

To the best of our knowledge, this is the first research to assess framework code samples.
We proposed a large scale empirical study to better understand how these code samples are maintained and used by developers.
By assessing 233 code samples provided by the worldwide frameworks Android and SpringBoot, we investigated aspects related to their source code, evolution, popularity, and client usage.
We reiterate the most interesting implications to support maintenance and usage of code samples:

\begin{itemize}
    
    \item Code samples should be simple and small to facilitate their reuse, as stated by guidelines and followed by the majority of the code samples of Android and SpringBoot.
    
    \item Code samples should provide working environments to ease their usage and rely on automated build and integration tool to improve quality.
    
    \item Code samples are not static and should evolve over time. Updates are commonly performed to follow recent framework versions, otherwise the code samples become out of date and less relevant to the clients.
    
    \item Code samples may benefit from scenarios where their developers also contribute to the framework itself. In this case, migration delay may be decreased.
    
    \item Clients of code samples may explore them via the cycle fork-change-learn. Indeed, a strong minority of the client developers do rely on this cycle when using code samples.
    
\end{itemize}

As future work, we plan to extend this research by assessing the code samples provided by other frameworks and written in other programming languages (\eg~Google Maps and Twitter APIs).
We also plan to analyze other metrics relevant to the samples, such as security and readability.
Finally, we plan to perform a survey with the creators and the clients of the code samples to better understand, from their point of view, major code sample limitations and benefits.

\section*{Acknowledgements}

This research is supported by CAPES and CNPq.

\bibliographystyle{IEEEtran}
\bibliography{main}

\end{document}